\documentclass[12pt,a4paper]{article}
\usepackage{epsfig}
\usepackage{amsfonts}

\def\beq{\begin{equation}}
\def\eeq{\end{equation}}
\def\beqa{\begin{eqnarray}}
\def\eeqa{\end{eqnarray}}

\def\Eqn#1{Eq.~(\ref{#1})}

\def\Eqns#1#2{Eqs.~(\ref{#1}) and~(\ref{#2})}

\def\Fig#1{Fig.~{\ref{#1}}}

\def\Figss#1#2{Figs.~{\ref{#1}}-{\ref{#2}}}

\def\Sec#1{Section~{\ref{#1}}}

\def\Tab#1{Table~\ref{#1}}

\newcommand\sss{\scriptscriptstyle}
\def\ord{{\cal O}}

\def\hs{\hat s}
\def\t{\hat t}
\def\lra{\leftrightarrow}
\def\cA{{\cal A}}
\def\yw{y_{\sss W}}
\def\mw{m_{\sss W}}

\def\spa#1.#2{\left\langle#1\,#2\right\rangle}
\def\spb#1.#2{\left[#1\,#2\right]}
\def\spaa#1.#2.#3{\langle\mskip-1mu{#1}
                  | #2 | {#3}\mskip-1mu\rangle}
\def\spbb#1.#2.#3{[\mskip-1mu{#1}
                  | #2 | {#3}\mskip-1mu]}
\def\spab#1.#2.#3{\langle\mskip-1mu{#1}
                  | #2 | {#3}\mskip-1mu\rangle}
\def\spba#1.#2.#3{\langle\mskip-1mu{#1}^+
                  | #2 | {#3}^+\mskip-1mu\rangle}

\def\feynsl#1{
  \setbox0=\hbox{/} \setbox1=\hbox{$#1$}
  \dimen0=\wd0 \advance\dimen0 by -\wd1 \divide\dimen0 by 2
  \ifdim\wd0>\wd1 \raise.15ex\copy0\kern-\wd0\kern\dimen0\copy1\kern\dimen0
  \else \kern-\dimen0\raise.15ex\copy0\kern-\dimen0\kern-\wd1\copy1\fi}

\newskip\humongous \humongous=0pt plus 1000pt minus 100pt

\newif\ifdtup

\makeatletter
\def\@eqnnum{\hbox{\reset@font\rm(\theequation)}}
\let\make@eqnnum=\@eqnnum %
\def\eqnum#1{\dec@eqnnum \global\def\make@eqnnum{\reset@font\rm(#1)}%
\def\@currentlabel{#1}%
}
\def\inc@eqnnum{\addtocounter{equation}{1}}
\def\dec@eqnnum{\addtocounter{equation}{-1}}
\@definecounter{equation}%
\@addtoreset{equation}{section} %
\def\theequation@prefix{{\thesection}.} %
\def\theequation{\theequation@prefix\arabic{equation}}%
\makeatother

\begin{document}
\topskip 1cm

\begin{titlepage}
\hspace*{\fill}\parbox[t]{4cm}{
IPPP/01/14\\
DCPT/01/28\\
DFTT 10/2001\\
ILL-(TH)-01-3\\ \today}
\vfill
\begin{center}
{\Large\bf $W$ Boson Production with Associated Jets\\[7pt]
at Large Rapidities}\\
\vspace{1.cm}

{J.~R.~Andersen$^1$, V. Del Duca$^2$, F. Maltoni$^3$ and W.J. Stirling$^1$}\\
\vspace{.2cm}
{$^1$\sl Institute for Particle Physics Phenomenology\\
University of Durham\\
Durham, DH1 3LE, U.K.}\\

\vspace{.2cm}
{$^2$\sl I.N.F.N., Sezione di Torino\\
via P. Giuria, 1 - 10125 Torino, Italy}\\

\vspace{.2cm}
{$^3$\sl Department of Physics \\
University of Illinois at Urbana-Champaign  \\
Urbana, IL\ \ 61801, USA}\\

\vspace{.5cm}

\begin{abstract}
We analyse $W$ boson production at hadron colliders in association with 
one or two jets,
both with the exact kinematics and in the high-energy limit. We argue
that the configurations that are kinematically
favoured tend to have the $W$ boson forward in rapidity.
Thus $W$ boson production in association with jets lends itself
naturally to extensions to the high-energy limit, which we examine
both at leading order and by resumming higher-order corrections through
the BFKL theory.
\end{abstract}

\end{center}
 \vfill

\end{titlepage}

\section{Introduction}

In recent years strong-interaction processes characterised by two
large and disparate energy scales, which are typically the squared
parton center-of-mass energy $\hs$ and the squared momentum transfer $\t$,
with $\hs\gg \t$, have been analysed.
These processes can be divided into two categories: $a)$ inclusive
processes, such as dijet production in hadron collisions at large rapidity
intervals~\cite{Abachi:1996et,Abbott:2000ai}, forward jet production in
DIS~\cite{Aid:1995we,Breitweg:1999ed,Adloff:1999fa}, and $\gamma^*\gamma^*$
collisions in double-tag events, $e^+\, e^- \to e^+\, e^- +$
hadrons~\cite{Acciarri:1999ix}; $b)$ diffractive processes,
such as dijet production with a rapidity gap between the tagging jets,
either in hadron collisions~\cite{Abachi:1994hb,Abachi:1996gz,Abe:1995de}
or in photoproduction~\cite{Derrick:1996pb}.

The interest in these processes stems from the possibility that their
description in terms of perturbative-QCD calculations at a fixed order in the
coupling constant $\alpha_s$ might not be adequate, and that a resummation to
all orders of $\alpha_s$ of large contributions of the type of
$\ln(\hs/|\t|)$, performed through the BFKL
equation~\cite{Kuraev:1977fs,Balitsky:1978ic}, might be needed.  An
additional motivation for the analysis of processes in the limit $\hs\gg \t$,
and in particular for dijet production in hadron collisions at large rapidity
intervals, inclusive or with a rapidity gap, is to use it as a test ground
for the production of a Higgs boson in association with jets at the LHC. A
Higgs boson is mainly produced via gluon fusion, $g\,g\to H$, mediated by a
top-quark loop.  If the Higgs-boson mass is above the threshold for
vector-boson production, the Higgs boson decays mostly into a pair of $W$ or
$Z$ bosons. The signal, though, is likely to be swamped by the $W\,W$, QCD
and $t\,\bar t$ backgrounds. A Higgs boson of such a mass is also produced in
$q\,q\to q\,q\,H$ via electroweak-boson fusion, $W\,W$ and $Z\,Z\to H$,
though at a smaller rate~\cite{Gunion:1995zu}.  However, this would have a
distinctive radiation pattern with a large gap in parton production in the
central rapidity region, because the outgoing quarks give rise to forward
jets in opposite directions~\cite{Cahn:1984ip,Cahn:1987zv}, with no colour
exchanged between the parent quarks that emit the weak
bosons~\cite{Dokshitzer:1987nc,Bjorken:1993er}. Accordingly, the topology of
the final state has been used to reduce the overwhelming $W\,W+2$-jet
background~\cite{Barger:1995zq}. In fact, requiring in $W\,W+2$-jet
production two forward jets in
opposite directions, which entails a large dijet invariant mass,
makes the parton sub-processes to be dominated by gluon exchange
in the crossed channel, with the W's produced forward in rapidity.

In this paper, we analyse forward $W$ production in association with jets as a
natural extension of dijet production at hadron colliders and forward-jet
production in DIS, and as a process that for large dijet invariant masses
shares the same dynamical features (i.e., gluon exchange in the crossed channel)
as $W\, W + 2$-jet production with forward jets, but is considerably simpler
to analyse. There are additional reasons to consider this process: firstly, it
could be experimentally easier to pick up forward $W$ bosons that decay
leptonically than forward jets; once a forward lepton has triggered the event,
one observes the jets that are associated to it, with no limitations on their
transverse energy. Conversely, in a pure jet sample one usually triggers the
event on a jet of relatively high transverse energy, thus the triggering jet
cannot be too forward. Secondly, $W$ production in association with jets lends
itself naturally to extensions to the high-energy limit, since it favours
configurations with a forward $W$ boson, as we shall see in \Sec{sec:w2jet}.
We limit our analysis to $W$-boson production, however we expect the same 
kinematical and dynamical considerations we make in this work to apply to
$Z$-boson production as well.

In \Sec{sec:wkin} we examine the exact leading-order inclusive rapidity
distributions for $W+1$-jet and $W+2$-jet production, for each parton
subprocess, as well as the rapidity distribution of the $W$ boson when
the rapidity interval between the two jets is large. In \Sec{sec:if} we
review the high-energy factorisation and the derivation of the impact
factor for jet production, and we calculate the impact factor for
$W+1$-jet production. In \Sec{sec:rate} we consider the rate for
$W+2$-jet production in several high-energy approximations, and
in \Sec{sec:bfklmc} we review the BFKL Monte Carlo event generator.
In \Sec{sec:numer} we analyse several distributions and candidate BFKL
observables, such as the production rate as a function of the rapidity
interval between the two jets, the azimuthal angle decorrelation and the
mean number of jets. Finally, in the last section we draw our conclusions.

\section{Kinematics of $W+1$-jet and $W+2$-jet Production}
\label{sec:wkin}

In this section we analyse in detail the kinematics of $W$ production in
association with one or two jets, and we show that in $p\,p$ colliders
asymmetric configurations with a forward $W$ boson are naturally favoured.\footnote{
Unless stated otherwise, we always understand $W$ to include both $W^+$ and $W^-$
production.} The results presented here have been obtained using tree-level
matrix elements generated by MADGRAPH~\cite{Stelzer:1994ta}.

\subsection{$W+1$-jet production}
\label{sec:wjet}

We consider the hadroproduction of a $W$ boson with an associated jet.
At leading order (LO), the parton subprocesses are $q\, \bar{q}\to W\, g$ and $q\, g\to W\, q$.
The momentum fractions of the incoming partons are given through
energy-momentum conservation by
\beqa
x_a &=& {|p_{j_\perp}|\over\sqrt{s} } e^{y_j} + {m_\perp\over\sqrt{s} } e^{\yw}\,,
\nonumber\\
x_b &=& {|p_{j_\perp}|\over\sqrt{s} } e^{-y_j} + {m_\perp\over\sqrt{s} }
e^{-\yw}\, ,\label{wjkin}
\eeqa
with $p_{j_\perp}$ the jet (and the $W$) transverse momentum  and $m_\perp = \sqrt{\mw^2 +
|p_{j_\perp}|^2}$ the $W$ transverse mass.

What are the typical distributions in $y_j$ and in $\yw$?
At proton-antiproton colliders, the subprocess
$q\, \bar{q}\to W\, g$ is leading; since the incoming quark and antiquark
are valence quarks and the up and the down quark distribution functions
have different shapes, this entails an asymmetry in the rapidity
distribution of $W^+$ versus $W^-$ bosons, both in fully inclusive
(Drell-Yan) $W$ boson production~\cite{QCD} and in $W+1$-jet production,
and accordingly a
large plateau for the rapidity distribution of the $W$ boson as a whole.

\begin{figure}[t]
\begin{center}
\epsfig{file=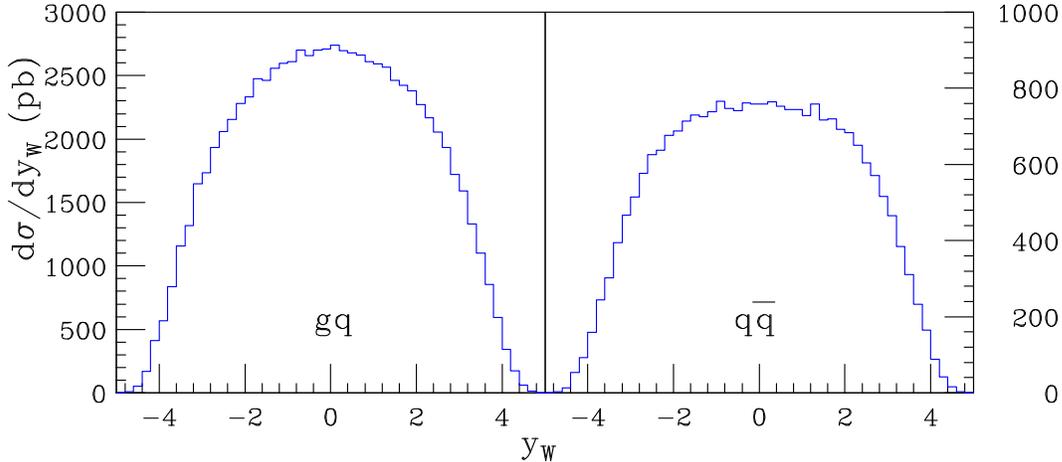,width=14cm}
\caption{Rapidity distributions of the $W$ boson for the subprocesses
$(a)$ $q\, g\to W\, q$ and $(b)$ $q\, \bar{q}\to W\, g$
at the LHC centre-of-mass energy $\sqrt{s} =$ 14 TeV and with
$p_{j_\perp{\rm min}} =$ 30 GeV.}
\label{fig:w1jetyw}
\end{center}
\end{figure}

Also at proton-proton colliders,
the $W$ boson may be produced abundantly in the forward rapidity region.
As in the $W^\pm$ rapidity asymmetry, the physical mechanism is the
difference in the shape of the p.d.f.'s of the incoming partons.
In fact, to be definite let us consider the subprocess $q\, g\to W\, q$,
which at proton-proton colliders is dominant, and
suppose that the incoming gluon enters from the negative-rapidity direction
while the quark enters from the positive-rapidity direction,
so we can identify $x_a$ as the gluon and $x_b$ as the quark momentum
fractions. The gluon distribution function is very steep, so it pays
off to have $x_a$ as small as possible, which can be achieved by taking
$\yw$ negative. That increases considerably the value of $x_b$, however,
because of the shape of the valence-quark distribution function, it
can be achieved without paying a high price.
In \Fig{fig:w1jetyw}, we consider the rapidity distribution of the
$W$ boson in $W+1$-jet production, broken up into its parton components.
The renormalisation and factorisation scales, $\mu_{\sss R}$ and
$\mu_{\sss F}$, are taken to be equal to $(|p_{j_\perp}| + m_\perp)/2$.
In \Figss{fig:w1jetyw}{fig:w2jetywy1y2} we have taken the $W$ mass to be
$\mw$ = 80.44 GeV, we have used the p.d.f.'s of the package MRST99cg
and evolved $\alpha_s$ accordingly~\cite{Martin:1998sq}. Applying the argument
above to both the gluon incoming directions for $q\, g\to W\, q$,
yields a broad rapidity distribution of the $W$ boson,
\Fig{fig:w1jetyw}a. The picture
above applies to the subprocess $q\, \bar{q}\to W\, g$ too,
\Fig{fig:w1jetyw}b, since the antiquark is in this case a sea quark.

\begin{figure}[t]
\begin{center}
\epsfig{file=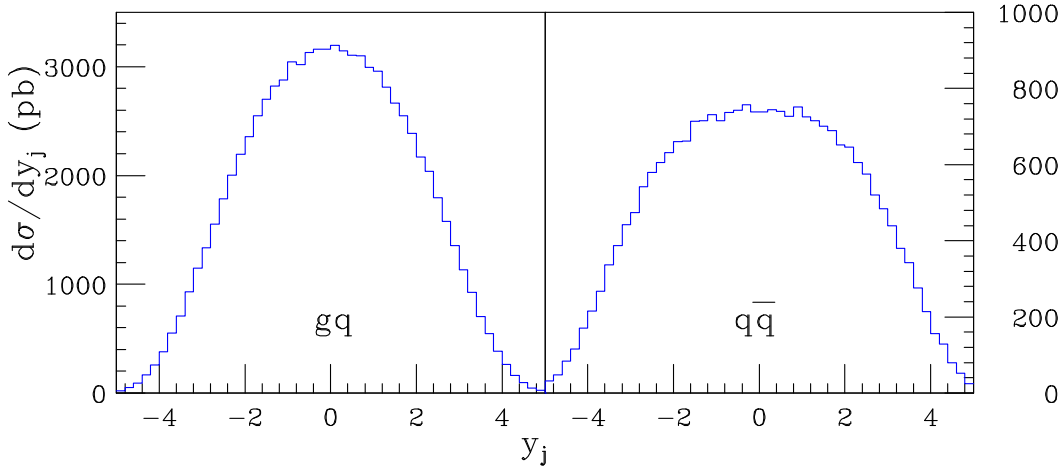,width=14cm}
\caption{Rapidity distributions of the jet for the subprocesses
$(a)$ $q\, g\to W\, q$ and $(b)$ $q\, \bar{q}\to W\, g$
at the LHC centre-of-mass energy $\sqrt{s} =$ 14 TeV and with
$p_{j_\perp{\rm min}} =$ 30 GeV.}
\label{fig:w1jety}
\end{center}
\end{figure}

In case the $W$ boson is produced forward in rapidity,
with which rapidity is the jet typically produced ?
If the jet is produced in the opposite hemisphere with respect to
the $W$ boson, the rapidity interval $|\yw - y_j|$ is large, however we know
that in this case $W+1$-jet production is strongly suppressed (at LO),
since its parton subprocesses can only have quark exchange in the crossed
channel, and thus the related production rate falls off with the parton
centre-of-mass energy $\hat s$.
Thus this configuration is dynamically disfavoured.
On the other hand, jet production in the same hemisphere as the $W$ boson or
centrally in rapidity
keeps $x_a$ small without substantially increasing $x_b$. However,
whether the jet is produced in the central region or in the same
hemisphere as the $W$ boson depends on the detailed shape of
the p.d.f.'s, namely on how large we can afford to make $x_b$
while keeping $x_a$ small. In \Fig{fig:w1jety} we plot the
rapidity distributions of the jet at LHC energies.

\subsection{$W+2$-jet production}
\label{sec:w2jet}

Let us consider the hadroproduction of a $W$ boson with two associated jets.
At LO the parton subprocesses are
\beqa
&(a)& g\, g\to W\, q\, \bar{q}\, ,\nonumber\\
&(b)& q\, \bar{q}\to W\, g\, g + W\, q\, {\bar q}\, ,\nonumber\\
&(c)& q\, q\to W\, q\, q\, ,\nonumber\\
&(d)& q\, g\to W\, q\, g\, .\label{subpro}
\eeqa
The momentum fractions of the incoming partons are given through
energy-momentum conservation by
\beqa
x_a &=& {|p_{j_{1\perp}}|\over\sqrt{s} } e^{y_{j_1}} +
{|p_{j_{2\perp}}|\over\sqrt{s} } e^{y_{j_2}} +
{m_\perp\over\sqrt{s} } e^{\yw}\,,
\nonumber\\
x_b &=& {|p_{j_{1\perp}}|\over\sqrt{s} } e^{-y_{j_1}} +
{|p_{j_{2\perp}}|\over\sqrt{s} } e^{-y_{j_2}} +
{m_\perp\over\sqrt{s} } e^{-\yw}\,
,\label{w2jkin}
\eeqa
with $p_{j_{1,2\perp}}$ the jet transverse momenta and
$m_\perp = \sqrt{\mw^2 + |p_{j_{1\perp}}+p_{j_{2\perp}}|^2}$ the
$W$ transverse mass. For the four subprocesses of \Eqn{subpro},
the total cross section for the production of a $W$ boson in
association with two jets is given in \Tab{tab:totx}.

\begin{table}
\begin{center}
\begin{tabular}{|l||c c|} \hline
subprocesses & {$\sigma(W^+)$ } & {$\sigma(W^-)$ }
\\ \hline\hline
$g\, g\to W\, q\, \bar{q}$
  & 170 & 170
\\ \hline
$q \bar{q}\to W\, g\, g + W\, q\, {\bar q}$
  & 580 &  400
\\ \hline
$q\, q\to W\, q\, q$
  & 400 &  300
\\ \hline
$q\, g\to W\, q\, g$
  & 3300 & 2200
\\ \hline
\end{tabular}
\caption{Total cross
sections (pb) for the production of $W^\pm$ boson in association with two jets
with transverse momentum $p_{j_{1,2\perp}} \ge$ 30 GeV and interjet distance
$R(j_1,j_2) = \sqrt{(y_{j_1}-y_{j_2})^2 + (\phi_{j_1}-\phi_{j_2})^2} \ge$ 0.4
on the rapidity-azimuthal angle plane.}
\label{tab:totx}
\end{center}
\end{table}

\begin{figure}[htb]
\begin{center}
\epsfig{file=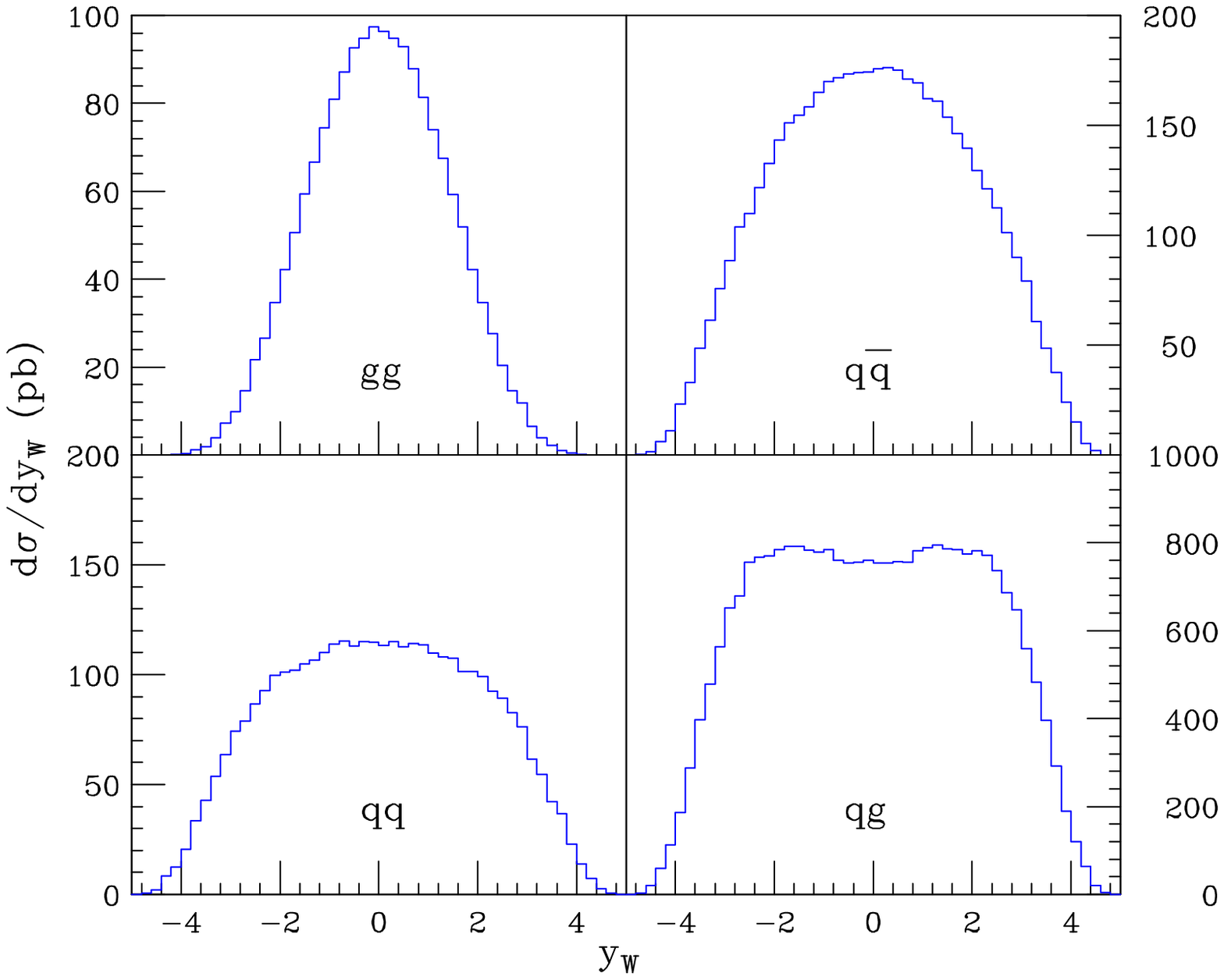,width=14cm}
\caption{Distributions in $\yw$ for the subprocesses of \Eqn{subpro} at
the LHC centre-of-mass energy $\sqrt{s} =$ 14 TeV and with
$p_{j_\perp{\rm min}}=$ 30 GeV.}
\label{fig:w2jetyw}
\end{center}
\end{figure}

\begin{figure}[htb]
\begin{center}
\epsfig{file=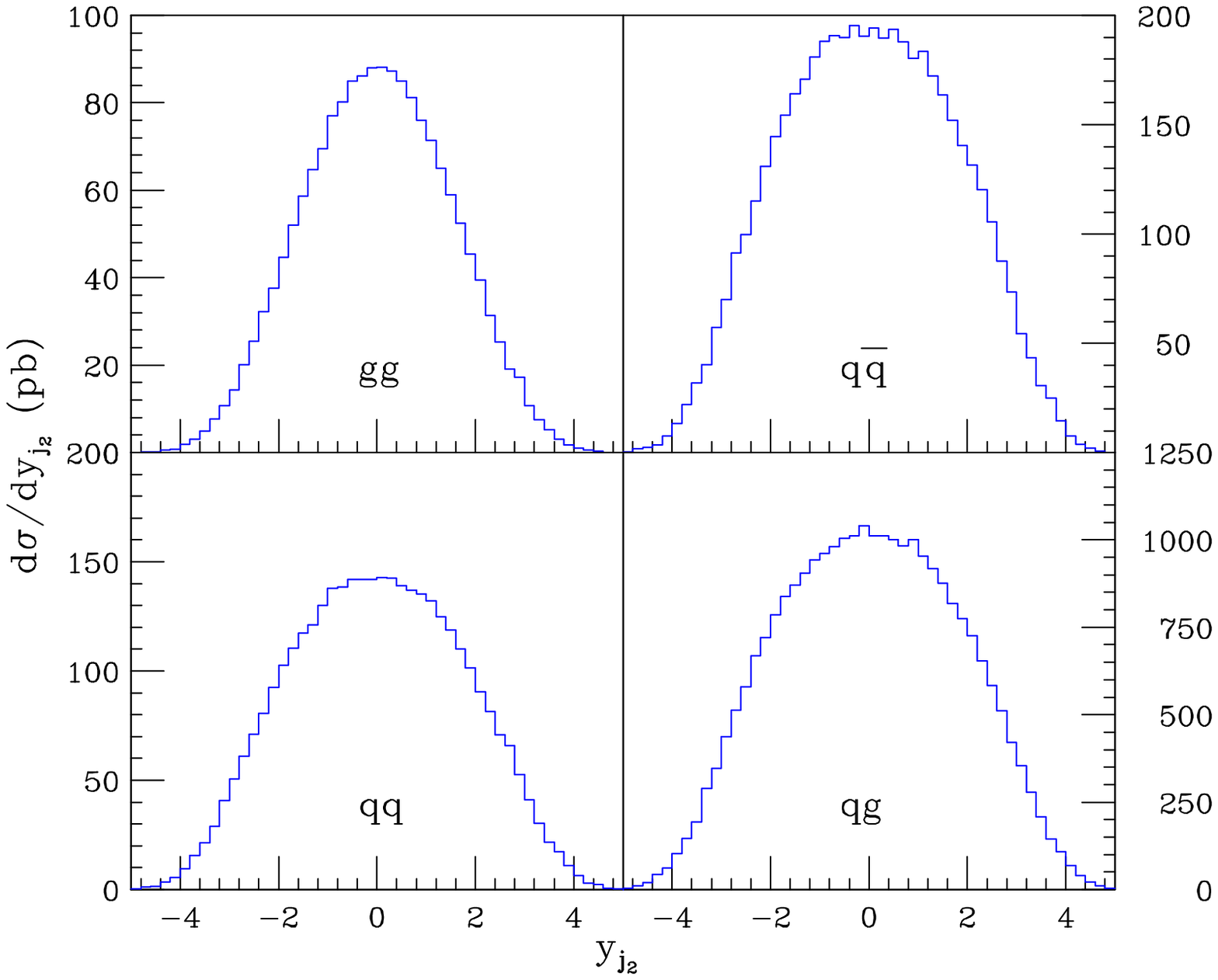,width=14cm}
\caption{Distributions in $y_{j_2}$, where $j_2$ is the jet that is closest to the
$W$, for the subprocesses of \Eqn{subpro} at
the LHC centre-of-mass energy $\sqrt{s} =$ 14 TeV and with
$p_{j_\perp{\rm min}} =$ 30 GeV.}
\label{fig:w2jety2}
\end{center}
\end{figure}

\begin{figure}[htb]
\begin{center}
\epsfig{file=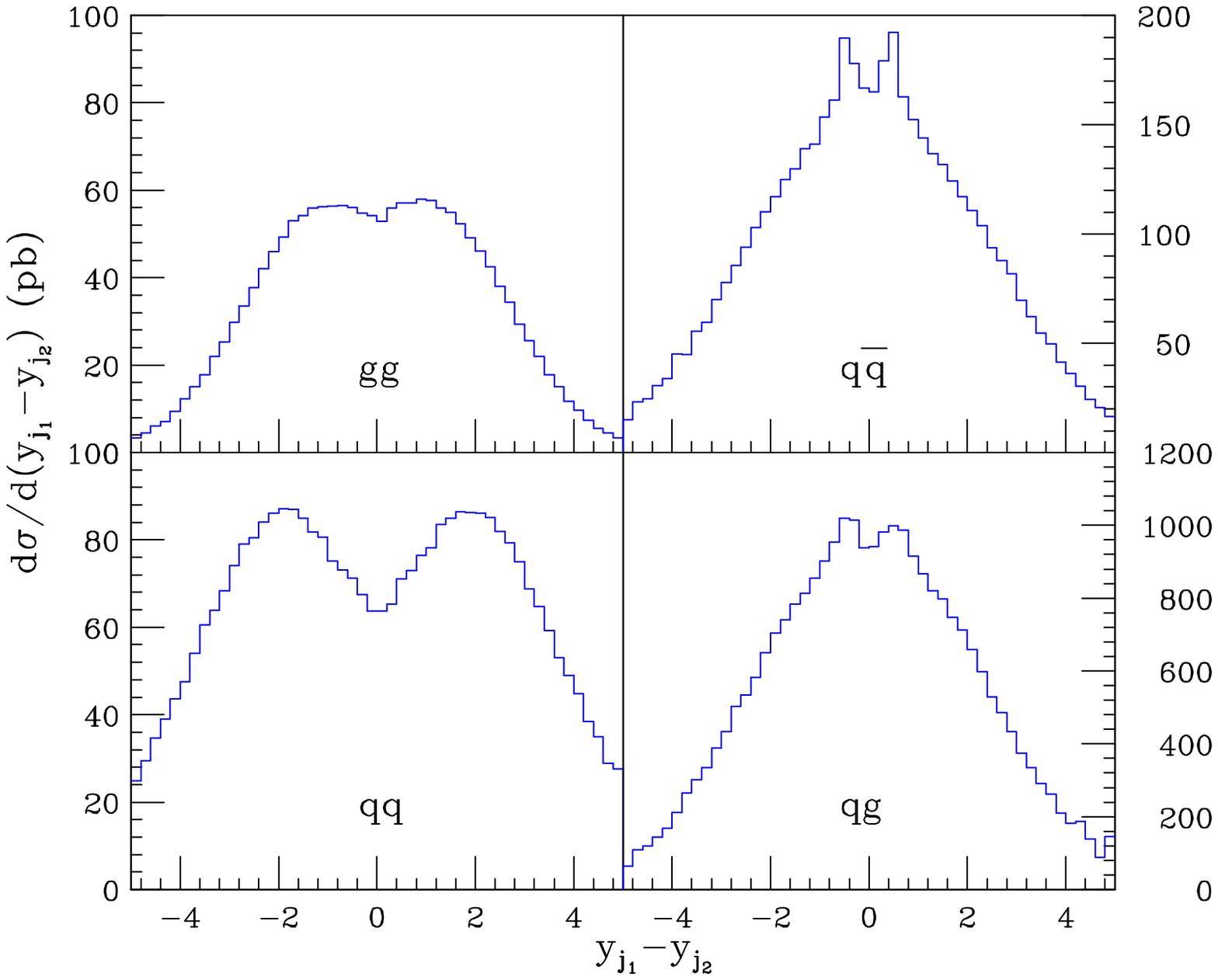,width=14cm}
\caption{Distributions in $y_{j_1}-y_{j_2}$ for the subprocesses of \Eqn{subpro} at
the LHC centre-of-mass energy $\sqrt{s} =$ 14 TeV and
with $p_{j_\perp{\rm min}} =$ 30 GeV.}
\label{fig:w2jety1y2}
\end{center}
\end{figure}

What are the typical rapidity distributions of the $W$ boson and of the
two jets? In \Figss{fig:w2jetyw}{fig:w2jetywy1y2} we plot the rapidity
distributions of the $W$ and of the two jets in $W+2$-jet production.
The renormalisation and factorisation scales, $\mu_{\sss R}$ and
$\mu_{\sss F}$, are taken
to be equal to $(|p_{j_{1\perp}}| + |p_{j_{2\perp}}| + m_\perp)/2$.
The subprocess $g\, g\to W\, q\, \bar{q}$ is perfectly symmetric,
thus the $W$ boson and the two jets are produced mostly in the central
rapidity region. However, in the other subprocesses
that is not the case: looking at the distributions in $\yw$
(\Fig{fig:w2jetyw})
we see that as
we move from $(a)$ to $(d)$ the $W$ boson tends to be produced
more and more forward in rapidity. Examining the distributions in $y_{j_2}$
(\Fig{fig:w2jety2}),
where $j_2$ is the jet that is closest to the $W$, we
see that this jet tends to follow the $W$ in rapidity. From the distributions
in $y_{j_1}-y_{j_2}$ (\Fig{fig:w2jety1y2}),
we see that in $(a)$ and $(b)$ jet 1 tends to be produced more centrally;
in $(d)$ it follows the $W$ boson and jet 2, thus emphasizing
the kinematical features already noted in $W+1$-jet production (the twin peaks
observed in \Fig{fig:w2jety1y2} in $(a)$, $(b)$ and $(d)$ are due to
requiring two jets with interjet distance $R(j_1,j_2)\ge$ 0.4);
finally in $(c)$ it tends to be produced far in rapidity from 
the $W$ boson and jet 2.

To understand how these configurations come about,
we consider $q\, g\to W\, q\, g$
and follow the analysis of \Sec{sec:wjet}, i.e.
we identify $x_a$ as the gluon and $x_b$ as the quark momentum
fractions. To make $x_a$ as small as possible at the price of
increasing $x_b$, the $W$ boson is produced forward
(\Fig{fig:w2jetyw}). Note that with
respect to \Sec{sec:wjet} this is made easier by the presence of two
jets, which let the $W$ boson have a transverse momentum as small as
kinematically possible: ultimately, when the jets are balanced in
transverse momentum, the $W$ transverse mass reduces to the mass,
$m_\perp \to m_{\sss W}$. In addition,
one jet, say $j_2$, is always linked to the $W$ boson via a quark
propagator as in $W+1$-jet production, so it tends to follow the $W$
in rapidity, as in \Fig{fig:w2jety2}, however the position
of the other jet is a dynamical feature peculiar of $W+2$-jet production:
thanks to the gluon exchanged in the crossed channel, that jet can be easily
separated in rapidity from the $W$ boson.
In $q \bar{q}\to W\, g\, g + W\, q\, {\bar q}$, the kinematical mechanism
is the same as in $q\, g\to W\, q\, g$ since the antiquark has a
sea quark p.d.f., however only $q \bar{q}\to W\, q\, {\bar q}$ can
have a gluon exchanged in the crossed channel.
For $g\, g\to W\, q\, \bar{q}$, which has equal p.d.f.'s for the incoming
particles and no gluon exchanged in the crossed channel, we obtain a 
central distribution, as expected. Note, however, that in \Fig{fig:w2jetyw}
and following the contribution of $g\, g\to W\, q\, \bar{q}$ to
$W+2$-jet production is quite small.
The $q\, q\to W\, q\, q$ channel is peculiar, since the largest contribution
comes from valence-quark distributions, which tend to have rather
large $x$'s. In addition, at the dynamical level it features
only diagrams with gluon exchange in the crossed channel.
Thus to make one $x$ large, it tends to have the $W$ boson and a jet 
slightly forward in rapidity, while to make the other $x$ large, it
has the second jet well forward (and opposite) in rapidity.

\begin{figure}[htb]
\begin{center}
\epsfig{file=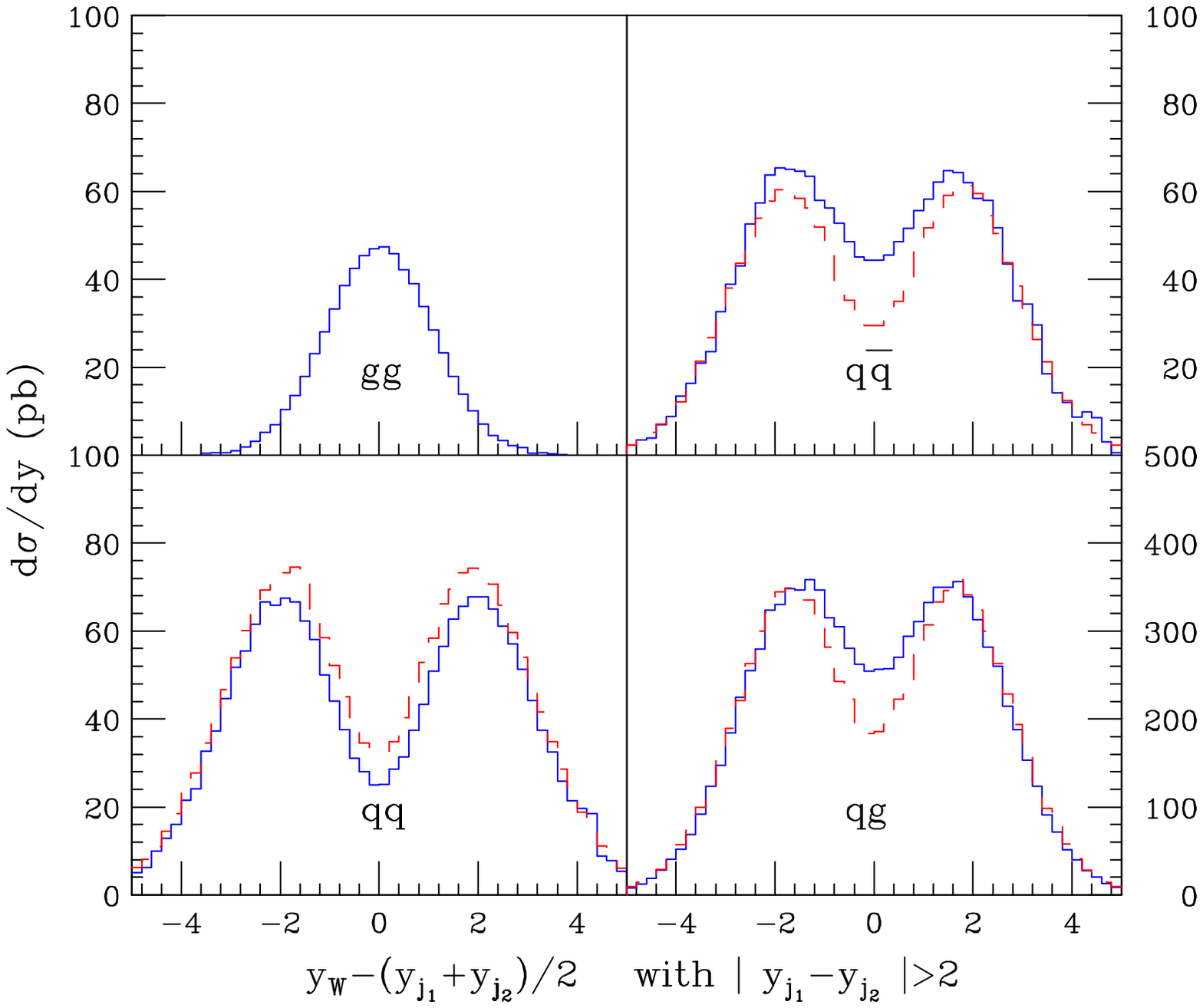,width=14cm}
\caption{Distributions of the rapidity of $W$ boson with respect to the
jet average,
$\yw-(y_{j_1}+y_{j_2})/2$ at $|y_{j_1}-y_{j_2}| \ge 2$ for the
subprocesses of \Eqn{subpro} at
the LHC centre-of-mass energy $\sqrt{s} =$ 14 TeV and
with $p_{j_\perp{\rm min}} =$ 30 GeV. The dashed line has been generated by
taking the amplitudes in the high-energy limit, as explained after
\Eqn{semiexact}.}
\label{fig:w2jetywy1y2}
\end{center}
\end{figure}

Next, we require that the two jets are produced with a sizeable rapidity
interval, $|y_{j_1}-y_{j_2}| \ge 2$, and look at the rapidity
distribution of the $W$ boson with respect to the jet average,
$\yw - (y_{j_1}+y_{j_2})/2$ (\Fig{fig:w2jetywy1y2}).
Now the requirement that the rapidity interval between the jets is large
makes the subprocesses with gluon exchange in the crossed channel
stand out even more, and the $W$ boson, which is linked to one of the
jets by quark exchange in the crossed channel, to follow that jet in rapidity.
This is stressed by the double peaks in $(b)$, $(c)$ and $(d)$.
Note that the dip between the peaks is maximal for $q\, q\to W\, q\, q$, 
which features only diagrams with gluon exchange in the crossed channel.
Conversely $g\, g\to W\, q\, \bar{q}$ yields the $W$ boson in the central 
rapidity region and approximately equidistant from the two jets, however it is
strongly suppressed since it can only have quark exchange in the
crossed channel.

The plots of \Fig{fig:w2jetywy1y2} are characterised by the dominance of the
subprocesses featuring gluon exchange in the crossed channel. The same feature
is exhibited by events where we select the jet, say $j_1$, that in rapidity is
furthest away from the $W$, require that $|\yw-y_{j_1}|\ge 2$, and examine the
distribution in $\yw-y_{j_2}$. Since in this case $j_2$ is always linked to the
$W$ boson by quark exchange, the distributions are all centered about zero. The
leading subprocesses factorise then naturally into two scattering centres: an
impact factor for $W+1$-jet production, and an impact factor for jet
production. The two impact factors are connected by the gluon exchanged in the
crossed channel. Accordingly, the dashed lines of Fig.~\ref{fig:w2jetywy1y2}
have been obtained by taking the high-energy limit of the amplitudes featuring
crossed-channel gluon exchange (see \Eqn{semiexact}). On these amplitudes we
can then insert the universal leading-logarithmic corrections of ${\cal
O}(\alpha_s^n\ln^n(\hs/|\t|))$, and resum them through the BFKL equation. The
impact factor for jet production is known, and we shall summarise its
derivation in \Sec{sec:jetif}. The impact factor for $W+1$-jet production is
derived in \Sec{sec:wif}. Next, we summarise jet production at hadron colliders
in the high-energy limit.

\section{Impact factors}
\label{sec:if}

In order to show how to extract the LO impact factor for $W+1$-jet
production, we use as a paradigm parton-parton scattering, and the derivation
of the LO impact factor for jet production.

\subsection{Dijet production at hadron colliders}
\label{sec:dijet}

In the high-energy limit $\hs\gg \t$, the BFKL theory assumes that any
scattering process is dominated by gluon exchange in the crossed channel,
which for a given scattering occurs at ${\cal O}(\alpha_s^2)$.
This constitutes the leading-order (LO) term of the BFKL resummation.
The corresponding QCD amplitude
factorizes into a gauge-invariant effective amplitude formed by
two scattering centers, the LO impact factors, connected by the gluon
exchanged in the crossed channel. The LO impact factors are characteristic
of the scattering process at hand. The BFKL equation resums then the universal
leading-logarithmic (LL) corrections, of ${\cal O}(\alpha_s^n\ln^n(\hs/|\t|))$,
to the gluon exchange in the crossed channel. These are obtained in the
limit of a strong rapidity ordering of the emitted gluon radiation,
\beq
y_1\gg y_2\gg \ldots \gg y_{n-1}\gg y_n\, .\label{mrk}
\eeq

For an arbitrary scattering, the LO term of the BFKL resummation
is contained in the higher-order terms
of the expansion in $\alpha_s$. For dijet production in hadron
collisions, the LO term of the BFKL resummation is already included in the
LO term of the expansion in $\alpha_s$. In this respect, dijet production
in hadron collisions at large rapidity intervals is the simplest process
to which to apply the BFKL resummation, and thus we shall use it as a paradigm.

Since the cross section for dijet production in the high-energy limit
is dominated by gluon exchange in the crossed channel,
the functional form of the QCD amplitudes for gluon-gluon, gluon-quark
or quark-quark scattering at LO is the
same; they differ only by the colour strength in the parton-production
vertices.  We can then write the cross section,
\begin{equation}
d\sigma = x_a f(x_a,\mu_F^2)\, x_b f(x_b,\mu_F^2)\, d\hat\sigma\,
,\label{xfact}
\end{equation}
in the following factorised
form~\cite{Mueller:1987ey,DelDuca:1994mn,Stirling:1994zs}
\begin{equation}
{d\sigma\over d^2 p_{a'_\perp} d^2 p_{b'_\perp} dy_{a'} dy_{b'}}\, =\,
x_a^0 f_{eff}(x_a^0,\mu_F^2)\, x_b^0 f_{eff}(x_b^0,\mu_F^2)\,
{d\hat\sigma_{gg}\over d^2 p_{a'_\perp} d^2 p_{b'_\perp}}\, ,\label{mrfac}
\end{equation}
where
\beq
x_a^0 = {|p_{a'_\perp}|\over\sqrt{s}} e^{y_{a'}}\,,\qquad \qquad
x_b^0 = {|p_{b'_\perp}|\over\sqrt{s}} e^{-y_{b'}}\,,\label{nkin0}
\eeq
are the parton momentum fractions in the
high-energy limit, $a'$ and $b'$ label the forward and backward outgoing jet,
respectively, and the effective parton distribution functions are
\begin{equation}
f_{eff}(x,\mu_F^2) = G(x,\mu_F^2) + {4\over 9}\sum_f
\left[Q_f(x,\mu_F^2) + \bar Q_f(x,\mu_F^2)\right], \label{effec}
\end{equation}
where the sum is over the quark flavours. In the high-energy limit,
the gluon-gluon scattering cross section becomes~\cite{Mueller:1987ey}
\begin{equation}
{d\hat\sigma_{gg}\over d^2 p_{a'_\perp} d^2 p_{b'_\perp}}\ =\
\biggl[{C_A\alpha_s\over |p_{a'_\perp}|^2}\biggr] \,
f(q_{a_\perp},q_{b_\perp},\Delta y) \,
\biggl[{C_A\alpha_s\over |p_{b'_\perp}|^2}\biggr] \ ,
\label{cross}
\end{equation}
with $\Delta y = y_{a'}-y_{b'}$ and $q_{i_\perp}$ the momenta transferred in
the $t$-channel, with $q_{a_\perp}=-p_{a'_\perp}$ and
$q_{b_\perp}=p_{b'_\perp}$, and where $C_{A}=N_{c}=3$.
The quantities in square brackets are proportional to the impact
factors for jet production. We shall analyse them in
\Sec{sec:jetif}. The function $f(q_{a_\perp},q_{b_\perp}, \Delta y)$ is the
Green's function associated with the gluon exchanged in the
crossed channel. It is process independent and given in the LL
approximation by the solution of the BFKL equation. Its analytic form is,
\beq
f(q_{a_\perp},q_{b_\perp},\Delta y)\, =
{1\over (2\pi)^2 |q_{a_\perp}| |q_{b_\perp}|}
\sum_{n=-\infty}^{\infty} e^{in\phi}\, \int_{-\infty}^{\infty} d\nu\,
e^{\omega(\nu,n)\Delta y}\, \left(|q_{a_\perp}|^2\over |q_{b_\perp}|^2
\right)^{i\nu}\, ,\label{solc}
\eeq
with $\phi$ the azimuthal angle between $q_a$ and $q_b$,
and $\omega(\nu,n)$ the eigenvalue of
the BFKL equation with maximum at $\omega(0,0)=4\ln{2}C_A\alpha_s/\pi$.
Thus the solution of the BFKL equation resums powers of $\Delta y$,
and rises with $\Delta y$ as $f(q_{a_\perp},q_{b_\perp},\Delta y) \sim
\exp(\omega(0,0)\Delta y)$.

\subsection{Impact factors for jet production}
\label{sec:jetif}

In the high-energy limit, $\ln(\hs/|\t|)\gg 1$,
the LO QCD amplitudes for gluon-gluon, gluon-quark
and quark-quark scattering all factorise into two impact factors for jet
production.  In order to determine explicitly the impact factors for
gluon-jet and for quark-jet production, we shall consider here two of
the subprocesses above.

The cross section for the scattering of two gluon into two gluons,
$g_a\,g_b\to g_{a'}\,g_{b'}$,
at LO in the high-energy limit,
$\hat s =x_a^0x_b^0 s \gg |\hat t|$, can be written as
\begin{eqnarray}
d\hat\sigma_{gg} &=& {1\over \hs}\,
\left[ {dy_{a'} d^2 p_{a'_\perp} \over 4\pi} \delta\left(\sqrt{s} x_a^0\, -\,
|p_{a'_\perp}| e^{y_{a'}}\right) \right]  \label{partxs}\\
&\times& \left[ {dy_{b'} d^2 p_{b'_\perp} \over 4\pi}
\delta\left(\sqrt{s} x_b^0\, -\,|p_{b'_\perp}| e^{-y_{b'}}\right) \right]
\delta^2(p_{a'_\perp} + p_{b'_\perp})\, |{\cal M}_{g\, g\to g\, g}|^2\,
,\nonumber
\end{eqnarray}
where $|{\cal M}_{g\, g\to g\, g}|^2$ is the squared tree amplitude, summed
(averaged) over final (initial) helicities and colours.
The amplitude for gluon-gluon scattering, $g_a\, g_b \rightarrow g_{a'}\,
g_{b'}$ with all external gluons outgoing, can be written
as~\cite{DelDuca:1995zy,DelDuca:1996km}
\begin{equation}
M^{aa'bb'\,}_{\nu_a\nu_{a'}\nu_{b'}\nu_b} = 2  \hs
\left[i g\, f^{aa'c}\, C^{gg}(p_a^{\nu_a},p_{a'}^{\nu_{a'}}) \right]
{1\over \t} \left[i g\, f^{bb'c}\, C^{gg}(p_b^{\nu_b},p_{b'}^{\nu_{b'}})
\right]\, ,\label{elas}
\end{equation}
where the $\nu$'s label the helicities and the LO vertices $g^*\, g
\rightarrow g$ are given by
\begin{equation}
C^{gg}(p_a^-, p_{a'}^+) = 1\,, \qquad C^{gg}(p_b^-, p_{b'}^+) =
{p_{b'_\perp}^* \over p_{b'_\perp}}\, ,\label{centrc}
\end{equation}
where we represent the transverse momentum $p_{\perp}$ on the complex plane,
$p_\perp =p_x+ip_y$. The functions $C$ transform
into their complex conjugates under helicity reversal,
$C^*(\{k^{\nu}\}) = C(\{k^{-\nu}\})$. The helicity-flip
function $C(p^+,p'^+)$ is subleading in the high-energy limit.
Thus each function, $C^{gg}$, has only the two helicity configurations of
Eq.~(\ref{centrc}) allowed in the high-energy limit.

We define the impact factor, $I(p_a,p_{a'})$ or $I(p_b,p_{b'})$, as the square
of each term in squared brackets in Eq.~(\ref{elas}),
summed (averaged) over final (initial) helicities and colours. Thus the
LO impact factor for a gluon jet is
\begin{eqnarray}
I^{g}(p_j,p_{j'}) &=& \frac{1}{2(N_c^2-1)}
\left[i g\, f^{jj'c}\, C^{gg}(p_j^{\nu_j},p_{j'}^{\nu_{j'}}) \right]
\left[- i g\, f^{jj'c'}\, [C^{gg}(p_j^{\nu_j},p_{j'}^{\nu_{j'}})]^* \right]
\nonumber\\ &=& g^2\, {C_A\over N_c^2-1}\, \delta^{cc'}\, ,\label{impgg}
\end{eqnarray}
with $j=a,b$ and with implicit sums over repeated indices.

{}From Eq.~(\ref{elas}) and (\ref{impgg}), the squared amplitude for
gluon-gluon scattering is
\begin{eqnarray}
|{\cal M}_{g\, g\to g\, g}|^2 &=& {4\hs^2\over \t^2}\,
I^{g}(p_a,p_{a'})\, I^{g}(p_b,p_{b'})\nonumber\\ &=&
{4 C_{A}^2\over N_c^2-1}\, g^4\, {\hs^2\over \t^2} = {9\over 2}\,
g^4\, {\hs^2\over \t^2}\, .\label{gg}
\end{eqnarray}
Analogously, the quark-gluon
$q_a\, g_b \rightarrow q_{a'}\, g_{b'}$ scattering amplitude in the
high-energy limit is~\cite{DelDuca:1996km}
\beqa
M^{aa'bb'\,}_{\nu_a\nu_b\nu_{b'}} = 2 {\hat s}
\left[g\, \lambda^c_{a' \bar a}\,
C^{\bar q q}(p_a^{\nu_a},p_{a'}^{-\nu_a}) \right] {1\over \t}
\left[i g\, f^{bb'c}\, C^{gg}(p_b^{\nu_b},p_{b'}^{\nu_{b'}}) \right]\,
,\label{elasqa}
\eeqa
with LO vertices $g^*\, q \rightarrow q$,
\beq
C^{\bar q q}(p_a^-,p_{a'}^+) = -i\, ;\qquad C^{\bar q q}(p_b^-,p_{b'}^+) = i\,
\left({p_{b'_\perp}^* \over p_{b'_\perp}}\right)^{1/2}\, .\label{cbqqm}
\eeq
Again, each function, $C^{\bar q q}$, has two helicity configurations allowed.
The LO impact factor for a quark jet is\footnote{We use the standard
normalization of the SU($N_c$) matrices, ${\rm tr} (\lambda^c \lambda^{c'}) =
\delta^{cc'} /2$.}
\begin{eqnarray}
I^{q}(p_a,p_{a'}) &=&{1 \over 2 N_c}
\left[g\, \lambda^c_{a' \bar a}\,
C^{\bar q q}(p_a^{\nu_a},p_{a'}^{-\nu_a}) \right]
\left[g\, \lambda^{c'}_{\bar a a'}\,
[C^{\bar q q}(p_a^{\nu_a},p_{a'}^{-\nu_a})]^* \right]
\nonumber\\ &=& {g^2\over 2N_c}\, \delta^{cc'}\, .\label{impqq}
\end{eqnarray}
The squared amplitude for quark-gluon scattering is then
\begin{eqnarray}
|{\cal M}_{q\, g\to q\, g}|^2 &=& {4\hs^2\over \t^2}\,
I^{q}(p_a,p_{a'})\, I^{g}(p_b,p_{b'})\nonumber\\ &=&
{2 C_{A}\over N_c}\, g^4\, {\hs^2\over \t^2} = 2\,
g^4\, {\hs^2\over \t^2}\, .\label{qq}
\end{eqnarray}
Since Eqs.(\ref{gg}) and (\ref{qq}) only differ by the colour factor 9/4,
to calculate the dijet production rate, \Eqn{mrfac}, it suffices to
consider one of them, such as gluon-gluon scattering, and include the others
through the effective p.d.f. in Eq.~(\ref{effec}). Using Eq.~(\ref{gg}) and
replacing $\t^2 \to |p_{a'_\perp}^2| |p_{b'_\perp}^2|$, the cross
section for gluon-gluon scattering (\ref{partxs}) becomes
\begin{equation}
{d\hat\sigma_{gg}^{0}\over d^2 p_{a'_\perp} d^2 p_{b'_\perp}}\ =\
\biggl[{C_A\alpha_s\over |p_{a'_\perp}|^2}\biggr] \,
{1\over2}\delta^{2}(p_{a'_\perp}+p_{b'_\perp}) \,
\biggl[{C_A\alpha_s\over |p_{b'_\perp}|^2}\biggr]\, .\label{cross0}
\end{equation}
At higher orders, powers of $\ln(\hat s/|\hat t|)$ arise, which can
be resummed to all orders in $\alpha_{s}\ln(\hat s/|\hat t|)$ through
the BFKL equation. The term between square brackets in \Eqn{cross0} is
the LO term of the BFKL resummation. Since in performing the resummation
the factorization formula (\ref{mrfac}) holds unchanged, to obtain the
resummed parton cross section it suffices to replace the LO term of the
BFKL ladder in \Eqn{cross0} with the full ladder, thus obtaining
\Eqn{cross}.

\subsection{The impact factor for $W+1$-jet production}
\label{sec:wif}

In $W+2$-jet production in the limit $|y_{j_1}-y_{j_2}|\gg 1$
(see \Sec{sec:w2jet}),
the parton subprocesses $q\, q\to W\, q\, q$ and $q\, g\to W\, q\, g$
and $q\, {\bar q}\to W\, q\, {\bar q}$ all feature gluon exchange in the
crossed channel. Thus the functional form of the corresponding
QCD amplitudes is the same. They differ only by the colour strength
in the impact factor for jet production, separated from the impact factor
for $W$ boson and the other jet by the gluon in the crossed channel.
It suffices then to consider only one of the subprocesses above.
We shall take $q\, g\to W\, q\, g$ and we shall suppose, for the sake of
clarity, that the $W$ boson is produced in the positive-rapidity hemisphere.
The subprocesses above factorise according to the kinematics
\beq
\yw \simeq y_q \gg y_g \,,\qquad\qquad |p_{\sss{W_\perp}}| \simeq |p_{q_\perp}|
\simeq |p_{g_\perp}| \,.\label{wkin}
\eeq
In the high-energy limit, the cross section for $q\, g\to W\, q\, g$
scattering is
\begin{eqnarray}
\label{sigmaW}
\lefteqn{
d\hat\sigma_{{\sss W} qg} = {1\over \hs}\,
\left[ {d\yw d^2 p_{\sss W} \over 4\pi (2\pi)^2}
{dy_q d^2 p_{q_\perp} \over 4\pi} \delta\left(\sqrt{s} x_a^0\, -\,
|p_{q_\perp}| e^{y_q} - m_\perp e^{\yw}\right) \right] } \label{whigh}\\
&\times&  \left[ {dy_g d^2 p_{g_\perp} \over 4\pi}
\delta\left(\sqrt{s} x_b^0\, -\, |p_{g_\perp}| e^{-y_g}\right) \right]
\delta^2(p_{{\sss W}_\perp} + p_{q_\perp} + p_{g_\perp})\,
|{\cal M}_{q\, g\to W\, q\, g}|^2\, ,\nonumber
\end{eqnarray}
with the $W$ transverse mass as in \Eqn{w2jkin}.
If we include the subsequent decay of the $W$ boson into a lepton pair,
the kinematics in the high-energy limit becomes
\beq
y_e \simeq y_\nu \simeq y_q \gg y_g \,,\qquad\qquad |p_{e_\perp}| \simeq
|p_{\nu_\perp}| \simeq |p_{q_\perp}| \simeq |p_{g_\perp}|\, .\label{enukin}
\eeq
The cross section for $q\, g\to q\, g\, (W\to) e\, \nu$ scattering can be
obtained from Eq.~(\ref{whigh}), replacing the $W$ boson with the lepton pair
and using the amplitudes calculated in Refs.~\cite{Gunion:1985vc,Kleiss:1985yh,Ellis:1985jg}. In the notation of
Ref.~\cite{Bern:1998sc} the colour decomposed amplitude is
\begin{eqnarray}
\lefteqn{ \cA_6(1_q,2,3,4_{\bar q};5_{\bar e},6_e) } \label{treew}\\
&=& g_W^2 g^2\, {\cal P}_W(s_{56})\, \sum_{\sigma\in S_{2}}
   (\lambda^{a_{\sigma(2)}} \lambda^{a_{\sigma(3)}})_{i_1}^{\bar i_4}\
    A_6(1_q,\sigma(2),\sigma(3),4_{\bar q};5_{\bar e},6_e)\, ,\nonumber
\end{eqnarray}
where legs $1,4$ are the $q \bar q$ pair, legs $2,3$ are the gluon legs,
and legs 5,6 are the lepton pair; $g_W$ is the weak coupling and
${\cal P}_W(s)$ the $W$ propagator
\begin{equation}
{\cal P}_W(s) = {1\over s - M_W^2 + i\, \Gamma_W\, M_W}\, ,
\end{equation}
where $M_W$ and $\Gamma_W$ are the mass and width of the $W$.
The colour ordered subamplitudes are
\begin{eqnarray}
\lefteqn{ A_6(1_q^-,2^-,3^+,4_{\bar q}^+,5_{\bar\ell}^+,6_\ell^-)
= {2\over s_{23} }\, \biggl[
   {\spb1.3 \spa2.1 \spb5.4 \spab6.{(1+2)}.3
       \over \spb2.1 t_{123} } } \nonumber\\
&-& {\spb4.3\spa2.4 \spa6.1 \spab2.{(3+4)}.5
       \over \spa4.3 t_{234} }
  - { \spab2.{(3+4)}.5 \, \spab6.{(1+2)}.3
       \over \spb2.1 \spa4.3 } \biggr]\,, \nonumber\\
\lefteqn{ A_6(1_q^-,2^+,3^-,4_{\bar q}^+,5_{\bar\ell}^+,6_\ell^-)
= {2\over s_{23} }\, \biggl[
- {\spa3.1^2\spb5.4 \spab6.{(1+3)}.2 \over \spa2.1 t_{123} }
} \nonumber\\
&+& {\spb4.2^2\spa6.1 \spab3.{(2+4)}.5 \over \spb4.3 t_{234} }
+ {\spa3.1\spb4.2\spa6.1\spb5.4 \over \spa2.1\spb4.3 }
   \biggr]\, ,\label{wsub}
\end{eqnarray}
with the spinor products defined in Appendix~\ref{sec:appa}.
The subamplitudes~(\ref{wsub}) are symmetric under the
exchange~\cite{Bern:1998sc}
\begin{equation}
1\lra 4\, ,\quad 2\lra 3\, ,\quad 5\lra 6\, ,\quad \langle ij\rangle\lra
[ji]\, .\label{flip}
\end{equation}
In Eq.~(\ref{wsub}) we have neglected the subamplitudes with
like-helicity gluons because they are subleading in the high-energy limit.

Next, we make the correspondence $p_4\equiv p_a$, $p_1\equiv p_{a'}$,
$p_2\equiv p_b$ and $p_3\equiv p_{b'}$ and according to \Eqn{wsub}
we always identify (5)6 as the (anti)lepton momentum.
The amplitude for $q_a\, g_b\to q_{a'}\, g_{b'}\, (W\to) e\, \nu$
scattering~(\ref{treew}) in the high-energy limit,
$y_q \simeq y_{e} \simeq y_{\nu} \gg y_{b'}$, is obtained by
computing the sub-amplitudes~(\ref{wsub}) in the corresponding
kinematics (Appendix~\ref{sec:appc})
\begin{eqnarray}
\lefteqn {\cA_{q_a\, g_b\to q_{a'}\, g_{b'}\, e\, \nu} } \label{wamp}\\
&=& 2  \hs
\left[g\, \lambda^c_{a' \bar a}\,
C^{\bar q q}(p_a^{\nu_a},p_q^{-\nu_a},
p_e,p_\nu,q) \right]
{1\over \t} \left[i g\, f^{bb'c}\,
C^{gg}(p_b^{\nu_b},p_{b'}^{\nu_{b'}}) \right]\, ,\nonumber
\end{eqnarray}
with $p_{b'}\equiv p_g$, and $q_\perp$ the momentum
transferred in the crossed channel, $q_\perp =p_{b'_\perp}$, and $\t \simeq
-|q_\perp|^2$, and the vertex $g^*\, d \to u\, e^-\, \bar{\nu}$
given by
\begin{eqnarray}
\lefteqn{ C^{\bar d \, u}(p_a^+,p_q^-,p_e,p_\nu,q) =
\frac{i g_{\sss W}\, {\cal P}_W(s_{\nu e}) }{\sqrt{(p^+_q+p^+_{\sss W})
{p^+_\nu}}} } \label{wdu}\\ &\times& \left( {\langle p_q p_e \rangle} \,
\frac{{p^+_\nu q_\perp^*} -  \left(p^+_q+p_{\sss W}^+\right) p_{\nu_\perp}^*}
{t_{abb'}}
+{\sqrt{\frac{p^+_q}{p^+_e}}}\,{p_{\nu_\perp}^*}
\frac{ p_{{\sss W}_\perp} p^+_e + p^+_q \,p_{e_\perp}}{t_{a'bb'}}
\right)\, ,\nonumber
\end{eqnarray}
with
\begin{eqnarray}
t_{a'bb'} &=& (p_a+p_{\sss W})^2 \simeq {- p^+_q p_W^- -
|p_{{\sss W}_\perp}|^2} \,,\nonumber \label{tinv}\\
t_{abb'} &=& (p_q+p_{\sss W})^2 \,,
\end{eqnarray}
and with $p_{\sss W} = p_e + p_\nu$. In the argument of the vertex in
the left-hand side of \Eqn{wdu} we do not write explicitly the helicity
of the lepton pair, since that is uniquely fixed by the helicity of the
quark pair and of the $W$ boson.
The vertex $g^*\, u \to d\, e^+\, \nu$ is obtained from
Eq.~(\ref{wdu}) by exchanging $(p_\nu \leftrightarrow p_e)$
\begin{equation}
C^{\bar u\,  d}(p_a^+,p_q^-,p_e,p_\nu,q)=
C^{\bar d \, u}(p_a^+,p_q^-,p_\nu,p_e,q)\, .\label{wud}
\end{equation}
Using the symmetry~(\ref{flip}) of the vertex~(\ref{wdu}),
we obtain the vertex $g^*\, \bar{d} \to \bar{u}\, e^+\, \nu$
\begin{equation}
C^{ d \, \bar u}(p_a^-,p_q^+,p_e,p_\nu,q) =
- \left[C^{\bar d \, u}(p_a^+,p_q^-,p_e,p_\nu,q)\right]^*\,
.\label{wdbaru}
\end{equation}
The vertex $g^*\, \bar{u} \to \bar{d}\, e^-\, \bar\nu$ is then
obtained from Eq.~(\ref{wdbaru}) by exchanging $(p_\nu \leftrightarrow p_e)$
\begin{equation}
C^{ u \,\bar d}(p_a^-,p_q^+,p_e,p_\nu,q) =
C^{ d \, \bar u}(p_a^-,p_q^+,p_\nu,p_e,q)\, .\label{wubard}
\end{equation}

The impact factor for $W+1$-jet production,
$I^{q {\sss W}}$, can be obtained by squaring any of the effective
vertices~(\ref{wdu}-\ref{wubard}) and by integrating out the lepton pair;
however, by using \Eqns{treew}{wsub} we have computed directly the squared
amplitude for $q\, g\to q\, g\, W$ scattering, and compared it to
Ref.~\cite{Mangano:1990gs}. Taking then the high-energy limit~(\ref{wkin}),
the squared amplitude
summed (averaged) over final (initial) colours and helicities, reduces to
\begin{equation}
|{\cal M}_{q\, g\to {\sss W}\, q\, g}|^2 =
\frac{4 \hat{s} ^2}{\hat{t} ^2} I^{q {\sss W}}(p_a,p_q,p_{\sss W},q)
I^{g}(p_b, p_{b'})\, ,\label{wfact}
\end{equation}
with
\beq
I^{q {\sss W}}(p_a,p_q,p_{\sss W},q)  =
-\frac{\delta^{cc'}}{2 N_c t_{abb'} t_{a'bb'} } g^2
\frac{g_{\sss W}^2}{2}
\left[ m_{\sss W}^2 \frac{t_{abb'}}{t_{a'bb'}}
\left( z + \frac{t_{a'bb'}}{t_{abb'}} \right)^2 - \hat{t}
\left( 1+ z^2\right) \right]\, ,\label{wimp}
\eeq
where we have defined the momentum fraction
\begin{equation}
z = \frac{p^+_q}{p^+_q+p_{\sss W}^+}\, .
\end{equation}
Using \Eqn{tinv} and $\t \simeq -|q_\perp|^2$, we can rewrite the
impact factor (\ref{wimp}) as,
\begin{eqnarray}
\lefteqn{I^{q {\sss W}}(p_a,p_q,p_{\sss W},q) } \label{wimp2}\\
&=& -\frac{\delta^{cc'}}{2 N_c t_{abb'} t_{a'bb'} } g^2
\frac{g_{\sss W}^2}{2}
\left[ m_{\sss W}^2 \frac{
\left(- z|q_\perp|^2 + |p_{q_\perp}|^2 - |p_{{\sss W}_\perp}|^2
 \right)^2}{t_{abb'}t_{a'bb'}} + |q_\perp|^2
\left( 1+ z^2\right) \right]\, .\nonumber
\end{eqnarray}
In the small $|q_\perp|$ limit, the jet opposite to the impact factor
for $W+1$-jet production becomes collinear, and the cross section obtained
from the squared amplitude
(\ref{wfact}) yields an infrared real correction.
Since the latter may have at most a logarithmic enhancement as
$|q_\perp|\to 0$, the squared amplitude (\ref{wfact}) cannot diverge more
rapidly than $1/|q_\perp|^2$. This entails that in the small $|q_\perp|$
limit, the impact factor (\ref{wimp2}) must be at least quadratic in
$|q_\perp|$, $I^{q {\sss W}} \sim \ord(|q_\perp|^2)$. Using
$q_\perp = - (p_{q_\perp} + p_{{\sss W}_\perp})$, it is immediate to see
that this is the case for \Eqn{wimp2}.
In addition, as $q_\perp\to 0$ we have an almost on-shell gluon scattering
with a quark, then $p_{q_\perp}\to - p_{{\sss W}_\perp}$ and averaging over
the azimuthal angle of $q_\perp$,  \Eqn{wimp2} becomes
\begin{eqnarray}
\lim_{q_\perp\to 0} I^{q {\sss W}}
&=& \frac{\delta^{cc'}}{2 N_c} g^2
\frac{g_{\sss W}^2}{2}
\left(
\frac{|q_\perp|\, z  (1-z) }
{ |p_{{\sss W}_\perp}|^2 + z m_{\sss W}^2 }
\right)^2
\left[ { (1+ z^2) \left( |p_{{\sss W}_\perp}|^4 + z^2 m_{\sss W}^4
\right)
+ 4 z^2 m_{\sss W}^2 |p_{{\sss W}_\perp}|^2 \over
z ( |p_{{\sss W}_\perp}|^2 + z m_{\sss W}^2 )^2}  \right]\,\nonumber\\
&=& 4 \delta^{cc'}
\left(
\frac{|q_\perp|\, z  (1-z) }
{ |p_{{\sss W}_\perp}|^2 + z m_{\sss W}^2 }
\right)^2  |{\cal M}_{q\, g\to {\sss W}\, q}|^2 \, ,\label{wimp3}
\end{eqnarray}
which explicitly shows that the impact factor is positive definite
and that it factorises into the squared amplitude
for $q\, g\to W\, q$ scattering~\cite{QCD}, as it should.

\section{The production rate for $W+2$ jets  }
\label{sec:rate}

In the collision of two hadrons $A$ and $B$, the differential production
rate of a $W$ boson with two associated jets is given in terms of the
rapidities and transverse momenta by
\beqa
\lefteqn{
{d\sigma\over d^2 p_{j_{1\perp}} d^2 p_{j_{2\perp}} d^2 p_{{\sss W}_\perp}
dy_{j_1} dy_{j_2} d\yw} } \nonumber\\ &=&
\sum_{ij}\, x_a f_{i/A}(x_a,\mu_F^2)\, x_b f_{j/B}(x_b,\mu_F^2)\,
{|{\cal M}_{ij}|^2\over 256\pi^5 {\hat s}^2}\,
\delta^2\left( p_{{\sss W}_\perp} + p_{j_{1\perp}} + p_{j_{2\perp}} \right)\,,
\label{xsec}
\eeqa
with parton momentum fractions (\ref{w2jkin}). In \Eqn{xsec} the dynamics
of the scattering are fully contained in the squared amplitude.

In the limit $|y_{j_1}-y_{j_2}|\gg 1$, as discussed in \Sec{sec:wif},
we can identify an
outgoing parton with a (anti)quark, while the other, that we called
$b'$ according to the notation of \Sec{sec:wif}, can either be a quark
or a gluon. The cross
section for $W+2$-jet production can be written in the factorised
form (\ref{mrfac}), by substituting Eq.~(\ref{wfact})
and using \Eqn{effec},
\beqa
\lefteqn{
{d\sigma\over d^2 p_{q_\perp} d^2 p_{b'_\perp} d^2 p_{{\sss W}_\perp}
dy_q dy_{b'} d\yw} }\nonumber\\ &=& \sum_i x_a^0 Q_i(x_a^0,\mu_F^2)\,
x_b^0 f_{eff}(x_b^0,\mu_F^2)
{I^{q {\sss W}} I^{g}\over 32\pi^5 |q_{a_\perp}|^2 |q_{b_\perp}|^2}
{\delta^2(q_{a_\perp} - q_{b_\perp})\over 2}
\, , \label{diffx}
\eeqa
where $q_{a_\perp} = -p_{q_\perp} - p_{{\sss W}_\perp}$ and
$q_{b_\perp} = p_{b'_\perp}$,
and where we have substituted $\t^2$ with $|q_{a_\perp}|^2 |q_{b_\perp}|^2$.
In the first p.d.f. the sum is over (anti)quark flavours, and the impact
factors are given in Eqs.~(\ref{impgg}) and (\ref{wimp}). The last
term is the LO term of the BFKL resummation. Thus, to obtain the
BFKL-resummed cross section we just need to replace
$\delta^2(q_{a_\perp} - q_{b_\perp})/2$ with
$f(q_{a_\perp},q_{b_\perp},\Delta y)$, as in \Eqn{solc}.

\begin{figure}[tb]
\begin{center}
\epsfig{file=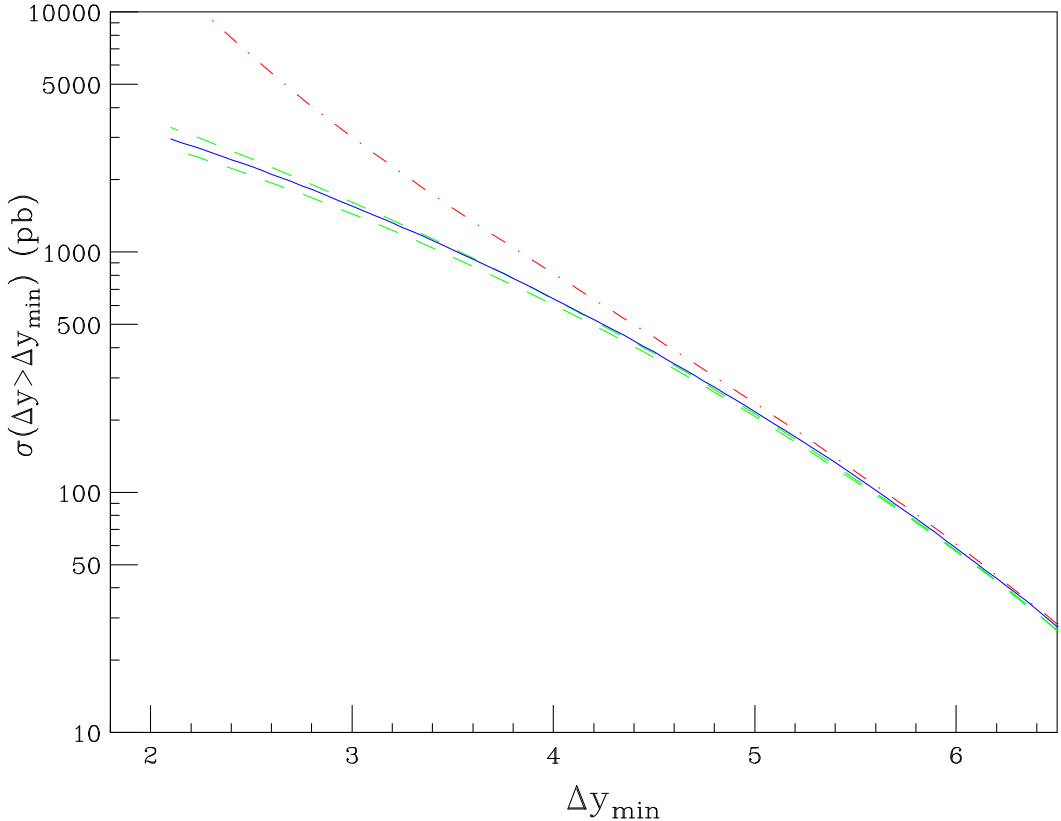,width=14cm}
\caption{The $W+2$-jet production rate as a function of the
rapidity interval between the jets $\Delta y = |y_{j_1}-y_{j_2}|$.
The solid curve is the exact production rate (\ref{xsec}); the dot-dashed
curve is the production rate in the high-energy limit (\ref{diffx});
the two dashed curves are given by the production rate (\ref{semiexact}),
with the two approximations for the dynamics mentioned in the text.}
\label{fig:modes}
\end{center}
\end{figure}

However, in \Eqn{diffx} energy and longitudinal momentum are not conserved.
The parton momentum fractions in the high-energy limit, $x_a^0$ and $x_b^0$,
given in \Eqn{sigmaW} underestimate the exact ones (\ref{w2jkin})
and accordingly the p.d.f.'s can be grossly overestimated. Thus for numerical
applications and for a comparison with experimental data,
it can be convenient to perform the high-energy limit only
on the dynamical part of \Eqn{xsec}, by writing the squared amplitude in the
factorised form (\ref{wfact}), while leaving the kinematics untouched.
This entails
\beqa
\lefteqn{
{d\sigma\over d^2 p_{q_\perp} d^2 p_{b'_\perp} d^2 p_{{\sss W}_\perp}
dy_q dy_{b'} d\yw} }\nonumber\\ &=& \sum_i x_a Q_i(x_a,\mu_F^2)\,
x_b f_{eff}(x_b,\mu_F^2) {1\over 32\pi^5}
\left[ { I^{q {\sss W}} I^{g}\over \hat{t}^2} \right]
{\delta^2(q_{a_\perp} - q_{b_\perp})\over 2}
\, .\label{semiexact}
\eeqa
For the invariants $\hat t$ and $t_{a'bb'}$,
implicit in the square brackets, two options are possible:\footnote{The
invariant $t_{abb'}$ is the same in the exact and high-energy kinematics.}
\begin{itemize}
\item[$(a)$] they are taken to be exact, namely
$\hat t = 2p_b\cdot p_{b'}$
and $t_{a'bb'} = (p_{a'}+p_{\sss W})^2$.
For instance, the dashed lines of Fig.~\ref{fig:w2jetywy1y2}
have been obtained from \Eqn{semiexact} with option $(a)$;
\item[$(b)$] $\hat t$ and $t_{a'bb'}$ are in the high-energy limit,
as defined below \Eqn{wamp} and in \Eqn{tinv} respectively.
\end{itemize}
Note that \Eqn{semiexact}, with the
two approximations for the dynamics above, and \Eqn{diffx}
have the same theoretical validity, however their numerics may be
rather different. In order to examine that in detail, in \Fig{fig:modes}
we consider $W+2$-jet production as a function of the
rapidity interval between the jets $\Delta y = |y_{j_1}-y_{j_2}|$. For
the renormalisation and the
factorisation scales we keep the same choice as in \Sec{sec:w2jet}.
The solid curve is the exact production rate~(\ref{xsec});
the dot-dashed curve is the production rate in the high-energy limit
(\ref{diffx}); the two dashed curves are given by the production rate
(\ref{semiexact}), with the two approximations listed above: $(b)$ is
the upper dashed curve, and $(a)$ is the lower one.
Note that the exact production rate is contained
between curves $(a)$ and $(b)$, with $(b)$ yielding the best numerical
approximation to the exact curve, while the high-energy limit on the LO
kinematics and dynamics (the dot-dashed curve) is rather
distant from the exact production rate, unless $\Delta y$ is quite large.
The range between curves $(a)$ and $(b)$ may be viewed as a
band of uncertainty on the high-energy limit at LO.

\section{The BFKL Monte Carlo}
\label{sec:bfklmc}

In a comparison with experimental data, it must be remembered that the LL
BFKL resummation makes some approximations which, even
though formally subleading, can be numerically important:
$a)$ The BFKL resummation is performed at fixed coupling constant,
thus any variation in the scale at which $\alpha_s$ is evaluated appears
in the next-to-leading-logarithmic (NLL) terms.
$b)$ Because of the strong
rapidity ordering any two-parton invariant mass is large. Thus there
are no collinear divergences in the LL resummation in the BFKL
ladder; jets are determined only at tree-level
and accordingly have no non-trivial structure.
$c)$ Finally, energy and longitudinal momentum are not conserved,
since the momentum fraction $x$ of the incoming parton is
reconstructed from the kinematic variables of the outgoing partons only, and
not including the radiation from the BFKL ladder. Therefore,
the BFKL theory can severely underestimate the exact value of the $x$'s,
and thus grossly overestimate the parton luminosities.
In fact, if a $W$ boson $+\ (n+2)$ partons are produced, we have
\beq
x_{a(b)} = {|p_{j_{1\perp}}|\over\sqrt{s} } e^{(-)y_{j_1}} +
{|p_{j_{2\perp}}|\over\sqrt{s} } e^{(-)y_{j_2}} +
{m_\perp\over\sqrt{s} } e^{(-)\yw}+
\sum_{i=1}^n {k_{i\perp}\over\sqrt{s}} e^{(-)y_i},\label{exkin}
\eeq
where the minus sign in the exponentials of the right-hand side applies
to the subscript $b$ on the left-hand side. In the BFKL theory, the LL
approximation and the kinematics (\ref{mrk}) imply that
in the determination of $x_a$ ($x_b$) only the
first (last) term in Eq.~(\ref{exkin}) is kept.
The terms neglected in Eq.(\ref{nkin0}) are formally subleading. However,
a comparison of three-parton production with the
exact kinematics (\ref{exkin}) to the truncation of the BFKL ladder to
${\cal O}(\alpha_s^3)$ shows that the LL approximation
entails sizable violations of energy-momentum
conservation~\cite{DelDuca:1995ng}.

Kinematic cuts and constraints like Eq.~(\ref{exkin}) can be implemented in
the BFKL framework by unfolding the BFKL integral equation, resulting in an
explicit sum over the number of emitted gluons. Each term in this sum is then
a phase space integral over the BFKL gluon phase space, which allows for a
BFKL Monte Carlo to be constructed~\cite{Orr:1997im,Schmidt:1997fg}.
Since each emitted BFKL
gluon enters the calculation with an explicit phase space integral, this
approach allows for the running of the coupling to be included into the BFKL
solution, and also for the gluon radiation to be taken into account in
Eq.~(\ref{exkin}).

The first step in this procedure is to transform the relevant Green's
function $f(\vec q_{a\perp},\vec q_{b\perp},\Delta y)$ of Eq.~(\ref{solc})
to moment space via
\begin{equation}
f(\vec q_{a\perp},\vec q_{b\perp},\Delta y) \ =\ \int {d\omega\over 2\pi i}\,
e^{\omega\Delta y}\,
f_{\omega}(\vec q_{a\perp},\vec q_{b\perp})\,.\label{moment}
\end{equation}
We can then write the BFKL equation as
\begin{equation}
\omega\, f_{\omega}(\vec q_{a\perp},\vec q_{b\perp})\, =
{1\over 2}\,\delta (\vec q_{a\perp}-\vec q_{b\perp})\, +\,
{\bar \alpha_s \over \pi}
\int {d^2\vec k_{\perp}\over k_{\perp}^2}\,
K(\vec q_{a\perp},\vec q_{b\perp},\vec k_{\perp})\, ,\label{bfklb}
\end{equation}
where the kernel $K$ is given by
\begin{equation}
K(\vec q_{a\perp},\vec q_{b\perp},\vec k_{\perp}) = f_{\omega}(\vec
q_{a\perp}+\vec k_{\perp},
\vec q_{b\perp}) - {q_{a\perp}^2\over k_{\perp}^2 +
(\vec q_{a\perp}+\vec k_{\perp})^2}\, f_{\omega}(\vec q_{a\perp},\vec q_{b\perp})
\, ,\label{kern}
\end{equation}
and $\bar\alpha_s=\alpha_s\frac {N_c} \pi$.
The first term in the kernel accounts for the emission of a real gluon of
transverse momentum $\vec k_{\perp}$ and the second term accounts for the
virtual radiative corrections.

We now separate the $\vec k_{\perp}$ integral in (\ref{bfklb}) into
`resolved' and `unresolved' contributions, according to whether $|\vec k_{\perp}|$ lie
above or below a small transverse energy scale $\mu$.  The scale $\mu$ is
assumed to be small compared to the other relevant scales in the problem
(such as the minimum transverse momentum, for instance).  The virtual and
unresolved contributions are then combined into a single, finite integral.
The BFKL equation becomes
\begin{eqnarray}
\omega\, f_{\omega}(\vec q_{a\perp},\vec q_{b\perp})\! &=&\!
{1\over 2}\,\delta (\vec q_{a\perp}-\vec q_{b\perp})\,
+\, {\bar\alpha_s\over \pi}
\int_{k_{\perp}^2 > \mu^2} {d^2\vec k_{\perp}\over k_{\perp}^2}\,
f_{\omega}(\vec q_{a\perp}+\vec k_{\perp},\vec q_{b\perp}) \nonumber \\
&+&\!\!\!\!  {\bar\alpha_s\over \pi} \int {d^2\vec k_{\perp}\over k_{\perp}^2}
\left[  f_{\omega}(\vec q_{a\perp}+\vec k_{\perp},\vec q_{b\perp})\, \theta(\mu^2 -k_{\perp}^2)\, - \, {q_{a\perp}^2 \, f_{\omega}(\vec q_{a\perp},\vec q_{b\perp})
 \over k_{\perp}^2 + (\vec q_{a\perp}+\vec k_{\perp})^2}
\right]\label{bfklbx}
\end{eqnarray}
The combined unresolved/virtual integral can be simplified by noting
that since $ k_{\perp}^2 \ll q_{a\perp}^2,q_{b\perp}^2$ by construction,
the $\vec k_{\perp}$ term in the argument of $f_{\omega}$ can be neglected,
giving
\begin{equation}
(\omega - \omega_0)\, f_{\omega}(\vec q_{a\perp},\vec q_{b\perp}) \, =\,
{1\over 2}\,\delta (\vec q_{a\perp}-\vec q_{b\perp})\,
+\, {\bar\alpha_s\over \pi}
\int_{k_{\perp}^2 > \mu^2} {d^2\vec k_{\perp}\over k_{\perp}^2}\,
f_{\omega}(\vec q_{a\perp}+\vec k_{\perp},\vec q_{b\perp}) \nonumber \\
\, ,\label{bfklcx}
\end{equation}
where
\begin{equation}
\omega_0  = {\bar\alpha_s\over \pi}
\int {d^2\vec k_{\perp}\over k_{\perp}^2}
\left[   \theta(\mu^2 -k_{\perp}^2)\, - \, {q_{a\perp}^2
 \over \vec k_{\perp}^2 + (\vec q_{a\perp}+\vec k_{\perp})^2}
\right]
\simeq {\bar\alpha_s}\, \ln\left( { \mu^2 \over q_{a\perp}^2 }   \right)
\, .    \label{eq:omega0}
\end{equation}
The virtual and unresolved contributions are now contained in $\omega_0$ and
we are left with an integral over resolved real gluons. Eq.~(\ref{bfklcx}) is
now solved iteratively, and performing the inverse transform we have
\begin{equation}
f(\vec q_{a\perp},\vec q_{b\perp},\Delta y)\, =  \,
  \sum_{n=0}^{\infty} f^{(n)}(\vec q_{a\perp},\vec q_{b\perp},\Delta y) \,,
\label{eq:b7}
\end{equation}
where
\begin{eqnarray}
f^{(0)}(\vec q_{a\perp},\vec q_{b\perp},\Delta y) &  =  &
\left[ \frac{\mu^2}{q_{a\perp}^2} \right]^{\bar\alpha_s\Delta y}\,
\,\frac{1}{2}\, \delta (\vec q_{a\perp}-\vec q_{b\perp} )\,,
\nonumber \\
f^{(n\geq 1)}(\vec q_{a\perp},\vec q_{b\perp},\Delta y) &  =  &
\left[ \frac{\mu^2}{q_{a\perp}^2} \right]^{\bar\alpha_s\Delta y}\,
\left\{ \prod_{i=1}^{n}  \int d^2 \vec k_{i\perp}\, dy_i \, {\cal F}_i \right\}
\,\frac{1}{2}\, \delta (\vec q_{a\perp}-\vec q_{b\perp} -
\sum_{i=1}^n \vec k_{i\perp})\,,
\nonumber \\
{\cal F}_i &=& \frac{\bar\alpha_s}{\pi k_{i\perp}^2}\,
\theta(k_{i\perp}^2 -\mu^2)\, \theta(y_{i-1}-y_i)\,
\left[ { (\vec q_{a\perp} +\sum_{j=1}^{i-1}\vec k_{j\perp}  )^2
 \over (\vec q_{a\perp} +\sum_{j=1}^{i}\vec k_{j\perp}  )^2 }
\right]^{\bar\alpha_s y_i}\,.
\label{eq:b8}
\end{eqnarray}
Thus the solution to the BFKL equation is recast in terms of phase space
integrals for resolved gluon emissions, with form factors representing the
net effect of unresolved and virtual emissions. In this way, each $f^{(n)}$
depends on the resolution parameter $\mu$, whereas the full sum $f$ does not.

\begin{figure}[!hbt]
\begin{center}
\epsfig{file=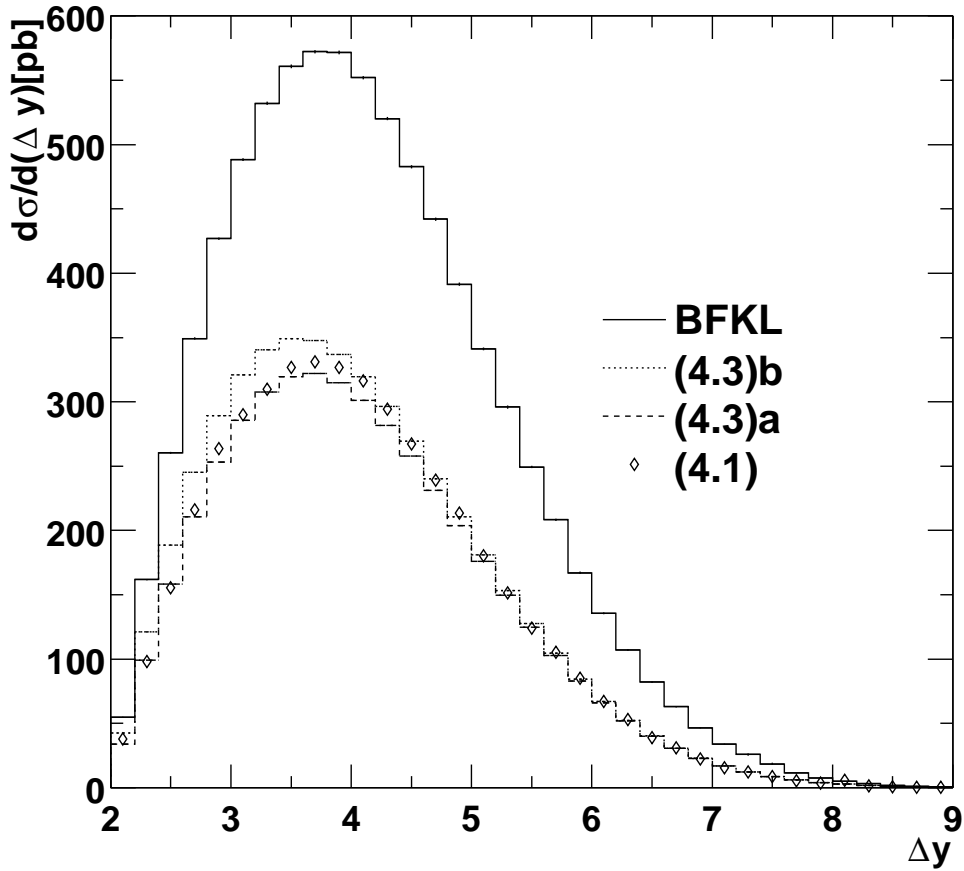,width=14cm}
\caption{The $W+2$-jet production rate as a function of the
rapidity interval between the jets $\Delta y = |y_{j_1}-y_{j_2}|$,
with acceptance cuts $y_{\sss W},\, y_{j_2} \ge 1$ and $y_{j_1} \le -1$,
or $y_{\sss W},\, y_{j_2} \le -1$ and $y_{j_1} \ge 1$.
The diamonds are the exact production rate (\ref{xsec});
the dashed curve is the production rate in the high-energy limit
(\ref{semiexact}) with option $(a)$; the dotted curve is the same with
option $(b)$; the solid curve includes the BFKL corrections.}
\label{fig:ywmy1}
\end{center}
\end{figure}

The derivation given above only applies for fixed coupling because we have
left $\alpha_s$ outside the integrals. The modifications necessary to account
for a running coupling $\alpha_s (k_{i\perp}^2)$ are
straightforward~\cite{Orr:1997im}. In the rest of this paper, we will however
discuss only the fixed coupling version of the BFKL Monte Carlo with the
coupling entering the BFKL equation set to $\alpha_s(p_{j_\perp{\rm min}}^2)$,
and with energy momentum conservation built in through Eq.~(\ref{exkin}). The
effects of including the BFKL gluon radiation in the Bjorken $x$'s are far
bigger than the effects of the running coupling, which amount to an
approximately 10\% effect in the cross section of Fig.~\ref{fig:ywmy1}.

\section{BFKL observables}
\label{sec:numer}

After having considered several approximations to the high-energy limit
and introduced the BFKL Monte Carlo as the tool that we shall use to
analyse the BFKL gluon radiation, we turn now to the analysis of the
effects of the BFKL radiation on some physical observables.
From \Eqn{solc} we see that in order to detect evidence of a BFKL-type
behaviour in a scattering process, we need to obtain $\Delta y$ as large as
possible.  In the context of dijet production this can be done by minimizing
the jet transverse energy, and maximizing the parton centre-of-mass energy
$\hat s$. Since $\hat s = x_a x_b s$, in a fixed-energy collider this is
achieved by increasing the parton momentum fractions $x_{a,b}$, and then
measuring the dijet production rate $d\sigma/ d\Delta y$. However, in dijet
production three effects hinder the characteristic growth of the BFKL ladder
(\ref{solc}) with respect to LO production:
\begin{itemize}
\item[] as the $x$'s grow the parton luminosities fall off, making it
difficult to disentangle the eventual BFKL-driven rise of the
parton cross section from the p.d.f.'s fall
off~\cite{DelDuca:1994mn,Stirling:1994zs};
\item[] the implementation of the exact $x$'s (\ref{exkin}) in the
BFKL Monte Carlo~\cite{Orr:1998hc}, rather than using $x^0_{a,b}$
(\ref{nkin0}) as prescribed by the high-energy limit, shifts the p.d.f.'s
toward smaller values, and thus further suppresses
the production rate. This effect is already present at
${\cal O}(\alpha_s^3)$~\cite{DelDuca:1995ng};
\item[] in dijet production both the tagged jets have typically the same
minimum transverse energy; at NLO, the dijet cross section as a function of the
difference $\cal D$ between the minimum transverse energies of the two
jets turns out to have a slope $d\sigma/d\cal D$ which is infinite at
${\cal D} = 0$~\cite{Frixione:1997ks,Andersen:2001kt}.
This hints to the presence of large logarithms of Sudakov type, which
can conceal the logarithms of type $\ln(\hat s/\hat t)$ characteristic
of the BFKL dynamics.\footnote{Logarithms of Sudakov type are contained
in the BFKL solution (\ref{solc}), however they lack the running of
$\alpha_s$ and they are not consistently resummed.}
\end{itemize}
The combination of these three effects changes drastically the shape
of dijet production $d\sigma/ d\Delta y$ as a function of the rapidity
interval between the tagged jets, showing a depletion~\cite{Orr:1998hc}
rather than the characteristic increase of the BFKL analytic solution.

In \Fig{fig:ywmy1} we consider $W+2$-jet production as a function of $\Delta
y$, and with acceptance cuts $y_{\sss W},\, y_{j_2} \ge 1$ and $y_{j_1} \le
-1$, or $y_{\sss W},\, y_{j_2} \le -1$ and $y_{j_1} \ge 1$. For all of the
curves of \Figss{fig:ywmy1}{fig:qperp}, we choose
$\mu_{\sss R1} = p_{j_{1\perp}}$ and $\mu_{\sss R2} = (p_{j_{2\perp}}
+ m_\perp)/2$ as renormalisation scales, and $\mu_{\sss F1} =
\mu_{\sss F2} = (|p_{j_{1\perp}}| + |p_{j_{2\perp}}| + m_\perp)/2$ as
factorisation scales. We justify the peculiar scale choices above as
follows: we note that our calculations are at LO (from the renormalisation
point of view), thus the scale choice is completely arbitrary,
as long as it is physically unambiguous. However, a uniform choice for
all of the curves in the same figure is required for a consistent
comparison between different approximations. In addition, in the
high-energy limit the impact factors for $W+1$-jet production on one side and
for jet production on the other can be viewed as two almost independent
scattering centres linked by a gluon exchanged in the crossed channel, thus
it makes sense to run $\alpha_s$ according to the scale set by each impact
factor. Accordingly, in the LO calculation $\alpha_s^2$ must be understood as
$\alpha_s(p_{j_{1\perp}}^2)\, \alpha_s\left( (p_{j_{2\perp}} + m_\perp)^2/4
\right)$. In the high-energy limit it is possible (and would make sense)
to choose the factorisation scales equal to the renormalisation scales,
however for the exact production rate this choice would not be physically
sensible since no high-energy factorisation is present, thus for the
factorisation scales we keep the same choice as in the previous figures. In
\Fig{fig:ywmy1} the diamonds represent the exact production rate
(\ref{xsec}); the dashed curve is the production rate in the high-energy limit
(\ref{semiexact}) with option $(a)$; the dotted curve is the same with
option $(b)$; the solid curve includes the BFKL corrections. In
\Figss{fig:ywmy1}{fig:qperp} we have computed the BFKL corrections using
\Eqn{semiexact} with option $(b)$. However, the particular option we choose
is immaterial since the uncertainty related to the choice of option in
\Eqn{semiexact} is much smaller than the uncertainties intrinsic to the BFKL
resummation, the latter being due to the leading-log approximation,
the choice of scale of $\alpha_s$ and the approximation on the
incoming parton momentum fractions. Note that the curve of \Fig{fig:ywmy1}
is both qualitatively and quantitatively different from $d\sigma/ d\Delta y$
in dijet production: the peak in \Fig{fig:ywmy1} is a striking
confirmation of the dominance of the configurations asymmetric in rapidity,
discussed in \Sec{sec:w2jet}.  In fact the symmetric acceptance cut
strongly penalises the asymmetric configurations when $\Delta y$
approaches its minimum value; since the asymmetric configurations dominate
the $W+2$-jet production rate, the effect is a strong depletion of the
latter. In addition, the BFKL ladder (solid curve), which includes
energy-momentum conservation~(\ref{exkin}), shows a substantial increase of
the cross section with respect to the LO analysis (dotted and dashed curves), as
opposed to a decrease in the dijet case.  To understand how this comes about,
we note that the presence of at least three particles in the final state
makes the threshold configurations, and thus the logarithms of Sudakov type,
much less compelling than in the dijet case.  Secondly, the implementation of
the kinematic constraint (\ref{exkin}) in the BFKL Monte Carlo, rather than
using $x^0_{a,b}$ (\ref{sigmaW}) in the high-energy limit, has a much lesser
impact than in the dijet case.  This is due to the fact that the valence
quark distribution in $q\, g\to q\, g\, W$ is much less sensitive to $x$
variations than the gluon distribution in $g\, g\to g\, g$.
\begin{figure}[tb]
\begin{center}
\epsfig{file=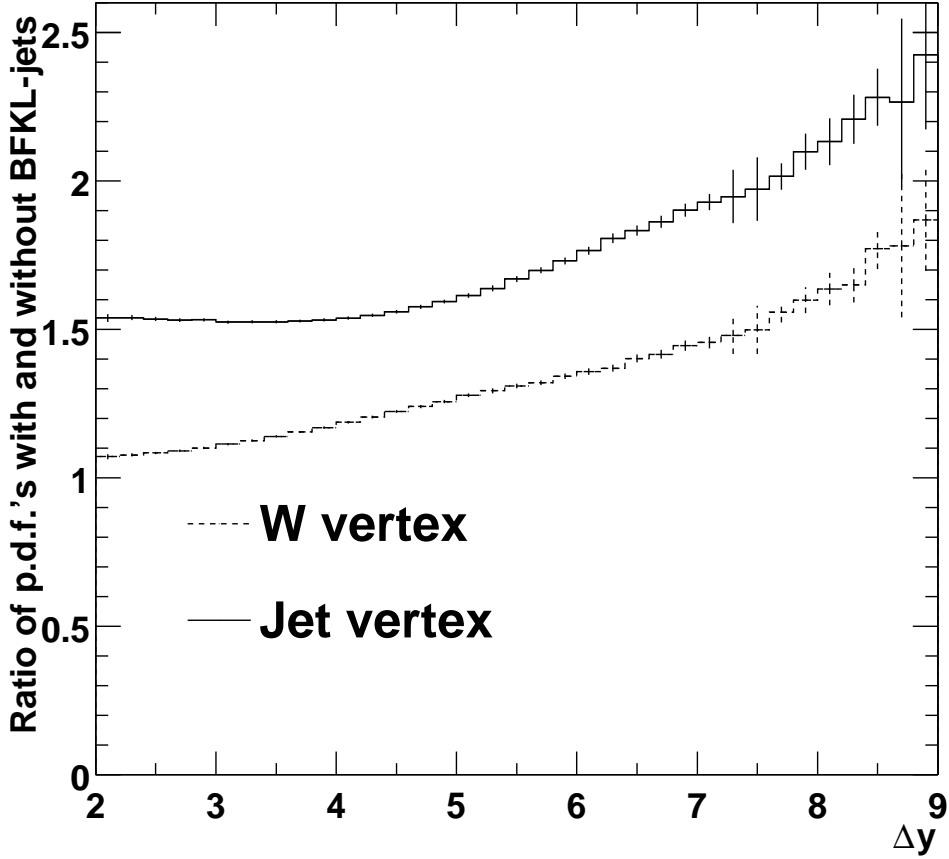,width=14cm}
\caption{The ratio $f(x^0,\mu_{\sss F}^2)/f(x,\mu_{\sss F}^2)$ of the
p.d.f.~as a function of the $x^0$'s (\ref{whigh})
in the high-energy limit versus the p.d.f.~as a function of the
exact $x$ (\ref{exkin}); the dashed-dotted curve is the ratio
$f(x_a^0,\mu_{\sss F}^2)/f(x_a,\mu_{\sss F}^2)$, and the solid
curve is the ratio
$f_{eff}(x_b^0,\mu_{\sss F}^2)/f_{eff}(x_b,\mu_{\sss F}^2)$.}
\label{fig:pdfract}
\end{center}
\end{figure}
To analyse this more precisely, we consider in \Fig{fig:pdfract}
the ratio $f(x^0,\mu_{\sss F}^2)/f(x,\mu_{\sss F}^2)$ of the p.d.f.~as a function of the $x^0$'s (\ref{whigh}) in the high-energy limit
versus the p.d.f.~as a function of the exact $x$ (\ref{exkin}).
The ratio is calculated for each event in the Monte Carlo as the ratio of the
p.d.f.~evaluated at $x^0$ compared to an evaluation at $x$, weighted with
the contribution of this event to the cross section according to
(\ref{semiexact}) with option $(b)$ and the BFKL ladder added. Finally, this
distribution is binned in $\Delta y$.
To be definite, since the high-energy factorisation entails
that each impact factor is
associated to one of the two incoming partons, we can term the ratio
$f(x_a^0,\mu_{\sss F}^2)/f(x_a,\mu_{\sss F}^2)$ as the one associated
to the impact factor for $W+1$-jet production, and the ratio
$f_{eff}(x_b^0,\mu_{\sss F}^2)/f_{eff}(x_b,\mu_{\sss F}^2)$ as the one
associated to the impact factor for jet production. As we see from
\Fig{fig:pdfract}, the solid curve is much farther away from 1 than the
dashed-dotted curve. Since the effective p.d.f.~is dominated by the
gluon distribution, this implies that the ratio
$f_{eff}(x_b^0,\mu_{\sss F}^2)/f_{eff}(x_b,\mu_{\sss F}^2)$
is much more sensitive to variations of the $x$'s than the ratio
$f(x_a^0,\mu_{\sss F}^2)/f(x_a,\mu_{\sss F}^2)$, which is made by
valence quark distributions. Accordingly, we obtain a smaller
depletion of the BFKL Monte Carlo prediction in $W+2$-jet production
as compared to dijet production. In addition, both the curves in
\Fig{fig:pdfract} rise as
$\Delta y$ grows. That entails that the BFKL radiation, which enters the
determination of the $x$'s in the denominator, yields as expected a
contribution which is growing with $\Delta y$.

\begin{figure}[tbh]
\begin{center}
\epsfig{file=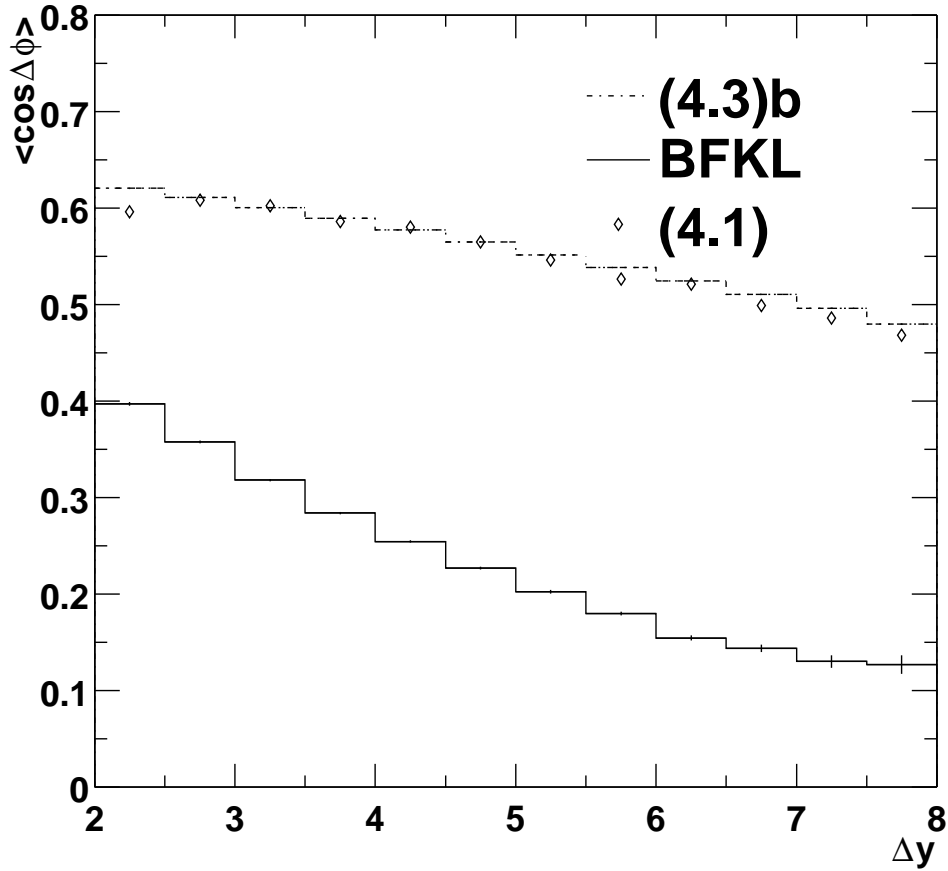,width=14cm}
\caption{The average azimuthal angle $\langle\cos \Delta\phi\rangle$, where
 $\Delta\phi=|\phi_{j_1}-\phi_{j_2}|-\pi$, as a function of the
rapidity interval between the jets $\Delta y = |y_{j_1}-y_{j_2}|$,
with acceptance cuts $y_{\sss W},\, y_{j_2} \ge 1$ and $y_{j_1} \le -1$,
or $y_{\sss W},\, y_{j_2} \le -1$ and $y_{j_1} \ge 1$.
The diamonds are the exact production rate (\ref{xsec});
the dashed-dotted curve is the production rate in the high-energy limit
(\ref{semiexact}) with option $(b)$; the solid curve includes the BFKL
corrections.}
\label{fig:avgcos}
\end{center}
\end{figure}

A variable that has been extensively studied as possibly sensitive to
BFKL effects is the azimuthal angle decorrelation
$\Delta\phi=|\phi_{j_1}-\phi_{j_2}|-\pi$
between the most forward and backward jets in inclusive dijet samples.
At LO the jets are supposed to be back to back, with a correlation which
is smeared by gluon radiation induced by parton showers and hadronization.
However, if we look at the correlation also as a function of $\Delta y$,
we expect the gluon radiation between the jets to
further blur the information on the mutual position in transverse momentum
space, and thus the decorrelation to grow with $\Delta y$.
From \Eqn{solc}, we see that the BFKL-induced gluon radiation
might account for that~\cite{DelDuca:1994mn,Stirling:1994zs,DelDuca:1995ng,
Orr:1997im,DelDuca:1995fx}. The decorrelation between the tagging jets has
been analysed, and indeed observed, by the D0 Collaboration in dijet
production at the Tevatron Collider~\cite{Abachi:1996et}.
However, the BFKL-induced radiation predicts a stronger decorrelation than
the data, even though the BFKL Monte Carlo~\cite{Orr:1997im} shows a much
more realistic azimuthal decorrelation than the BFKL analytic solution.
The data are correctly reproduced by the HERWIG Monte Carlo
generator~\cite{Marchesini:1988cf,Knowles:1988vs,Marchesini:1992ch},
which includes parton showers and hadronization. This suggests that
the azimuthal angle decorrelation $\Delta\phi$, picking up
preferentially configurations where the tagged jets are back to back,
is sensitive to threshold configurations, and thus to logarithms
of Sudakov type, even more than it is in the inclusive dijet production rate
$d\sigma/ d\Delta y$~\cite{Andersen:2001kt,Dokshitzer:1997uz}.
However, as discussed in the paragraph above, in $W+2$-jet production
we expect the logarithms of Sudakov type to play a much less significant
role. Thus,
in analogy with dijet production, in \Fig{fig:avgcos} we consider the
average azimuthal angle $\langle\cos\Delta\phi \rangle$ as a function of the
rapidity interval between the jets $\Delta y$.
The acceptance cuts are the same as for \Fig{fig:ywmy1}.
The diamonds are the exact production rate (\ref{xsec});
the dashed-dotted curve is the production rate in the high-energy limit
(\ref{semiexact}) with option $(b)$; the solid curve includes the BFKL
corrections. The average azimuthal angle being defined as a ratio of
production rates is much less sensitive to scale variations than the
curves of \Fig{fig:ywmy1}.

\begin{figure}[tbh]
\begin{center}
\epsfig{file=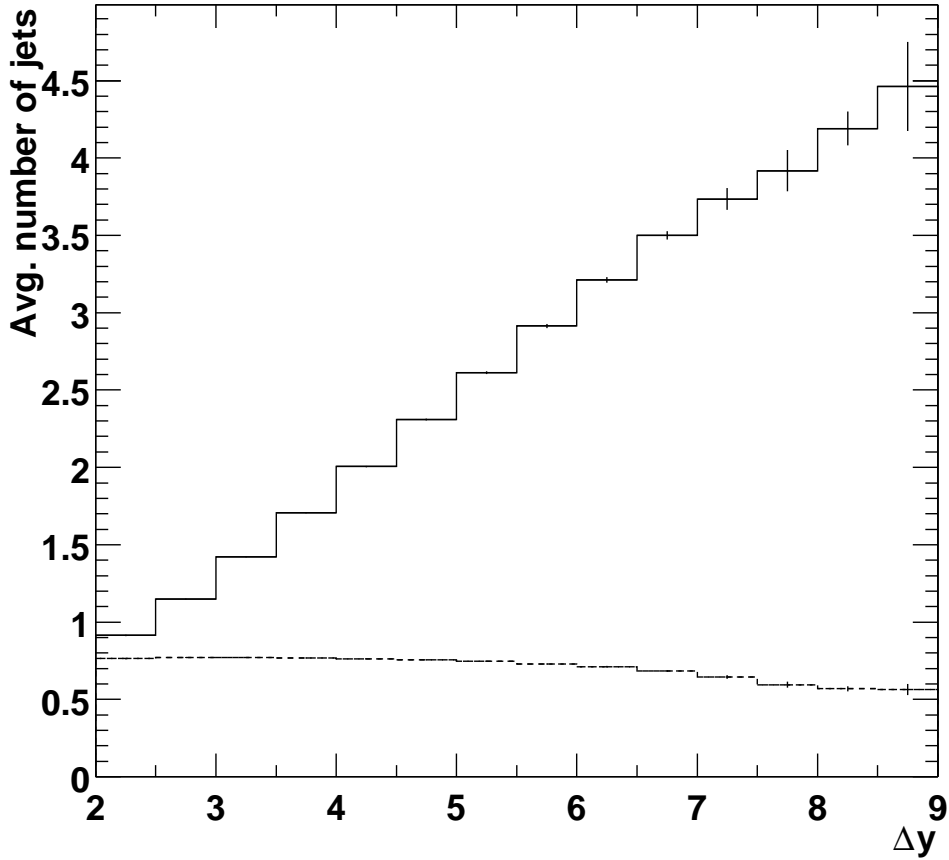,width=14cm}
\caption{Solid: the BFKL prediction for the mean number of jets
  $\langle n \rangle$ with $p_{\perp} > p_{j\perp \rm min}=30$~GeV as a
  function of the rapidity interval between the jets $\Delta y =
  |y_{j_1}-y_{j_2}|$. Dotted: the same in the rapidity range $-1 < y_j < 1$.}
\label{fig:avgjet}
\end{center}
\end{figure}

In \Fig{fig:avgjet} we plot the BFKL prediction for the mean number of jets
$\langle n\rangle$  with $p_{\perp} > p_{j\perp \rm min}=30$~GeV, emitted
by the BFKL ladder as a function of the rapidity interval between the jets
$\Delta y$, and the BFKL prediction for the same variable in the rapidity
range $-1 < y_j < 1$. We see that the mean number of jets rises
approximately linearly with $\Delta y$, and accordingly that the mean
number of jets in the rapidity range $-1 < y_j < 1$ stays constant. We can
crudely understand this, by noting that for a very large $\Delta y$ the
cross section from \Eqns{solc}{diffx} behaves like
\beq
\sigma|_{\Delta y} \sim e^{\omega(0,0)\Delta y} = \sum_{n=0}^\infty
{(\omega(0,0)\Delta y)^n \over n!}\,,
\eeq
with $\omega(0,0)=4\ln{2}C_A\alpha_s/\pi$, and a power of
$\alpha_s$ for each real correction to, and therefore for each
emitted gluon from, the BFKL ladder. Up to corrections of type
$\ln(p_{\perp}/p_{j\perp \rm min})$ \cite{Schmidt:1997fg},
the mean number of jets emitted by the BFKL ladder is then
\beq
\langle n\rangle = {(n \sigma)|_{\Delta y}\over \sigma|_{\Delta y} } \simeq
\omega(0,0)\Delta y \,.
\eeq
For $\alpha_s(p_{j\perp \rm min}^2)$ with $p_{j\perp \rm min}=30$ GeV, this
yields typically a jet each second unit of rapidity, which is in rough
agreement with \Fig{fig:avgjet}.

\begin{figure}[tbh]
\begin{center}
\epsfig{file=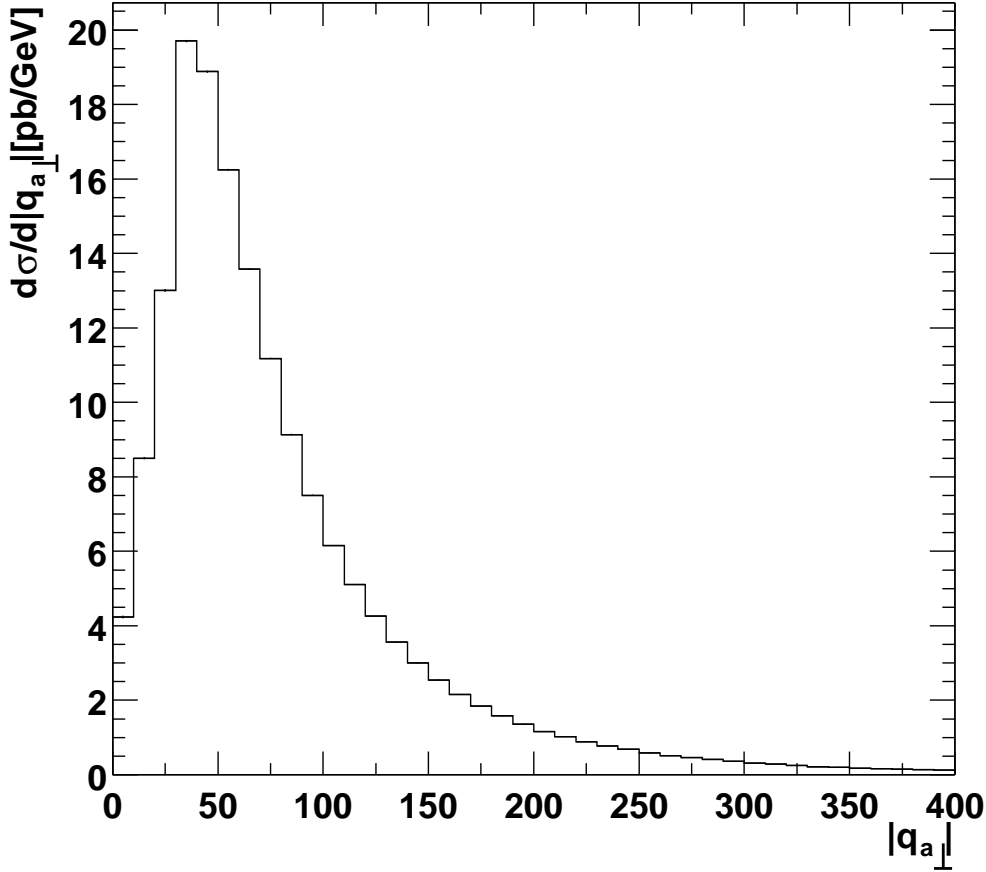,width=14cm}
\caption{The $W+2$-jet production rate, including BFKL corrections,
as a function of the transverse momentum $q_{a_\perp}$, with
$q_{a_\perp} = - (p_{q_\perp} + p_{{\sss W}_\perp})$.}
\label{fig:qperp}
\end{center}
\end{figure}

Finally, in \Fig{fig:qperp} we consider $W+2$-jet production as a function of
the transverse momentum $q_{a_\perp} = - (p_{q_\perp} + p_{{\sss W}_\perp})$
exiting from the impact factor $I^{q {\sss W}}$ for $W+1$-jet production.
At LO, $q_{a_\perp} = q_{b_\perp} = p_{b'_\perp}$, thus $q_{a_\perp}$ is
bound to be equal to the transverse momentum of the jet opposite to the
impact factor $I^{q {\sss W}}$ (and thus to be always larger than 30 GeV).
In presence of the gluon radiation of the BFKL ladder, this is
not longer true, and $q_{a_\perp}$ is allowed to go to zero. However,
a simple power-counting argument shows that the production rate is
finite as $q_{a_\perp} \to 0$. In fact from \Eqn{wimp2} we know that
$I^{q {\sss W}} \sim \ord(|q_{a_\perp}|^2)$. Substituting it, and the ladder
(\ref{solc}) which goes like $\ord(1/|q_{a_\perp}|)$, in \Eqn{diffx}, we
see that as far as the behaviour in $q_{a_\perp}$ is concerned,
\beq
{d\sigma\over dq_{a_\perp}^2} \sim {q_{a_\perp}^2\over q_{a_\perp}^3}
\delta^2(q_{a_\perp} + p_{q_\perp} + p_{{\sss W}_\perp})\, ,\label{powerc}
\eeq
and therefore the distribution $d\sigma/ dq_{a_\perp}$ is finite as
$q_{a_\perp} \to 0$, in agreement with \Fig{fig:qperp}.

\section{Conclusions}
\label{sec:conc}

In \Sec{sec:wkin} we have examined the exact LO inclusive rapidity
distribution for $W+1$-jet and $W+2$-jet production; as for the latter,
we have seen
that the dominant parton sub-process $q\, g\to q\, g\, W$ produces
a great deal of $W$ bosons forward in rapidity. This is due to the
different shape of the p.d.f.'s of the incoming quark and gluon, and
to gluon exchange in the crossed
channel which loosens the bound between the $W$ boson and a jet on one
(rapidity) side, and the other jet on the other side.
In \Sec{sec:if} we have derived the impact factor for $W+1$-jet production,
both as a function of the $W$-boson momentum and including the leptonic
decay of the $W$ boson.
In \Sec{sec:rate} we have compared several high-energy approximations
at LO to the exact production rate. The range between the most extreme
high-energy approximations may be considered as the theoretical
uncertainty on the high-energy limit at LO.

In \Sec{sec:numer} we have considered some BFKL footprints, most notably
the rate $d\sigma /d\Delta y$ and the azimuthal angle decorrelation
$d\sigma /d\Delta\phi$ as functions of the rapidity
interval $\Delta y$ between two tagged jets. These observables had already
been considered in inclusive dijet production, however because of the
dominance of the configurations asymmetric in rapidity and the presence
of at least three particles in the final state, which makes threshold
configurations less relevant, in $W+2$-jet production $d\sigma /d\Delta y$
and $d\sigma /\Delta\phi$ take on a completely new light.
In addition, we have considered the mean number of jets, which as expected
rises approximately linearly with $\Delta y$. Finally, we have computed
the transverse momentum distribution of the impact factor for
$W+1$-jet production. At LO this is bound from below by momentum
conservation at the minimum transverse energy of the jet opposite to
the $W+1$-jet configuration, but with additional gluon radiation it is
allowed to reach zero, where  the distribution is finite.

Finally, we note that one of the leading contributions to the
$WW+2$-jet production rate in the high-energy limit is obtained by
convoluting two impact factors
for $W+1$-jet production with a gluon exchanged in the crossed channel.
The analysis of this process, in the exact case, in the high-energy limit
and with BFKL corrections is left for future investigations~\cite{adms}.


\section*{Acknowledgments}

We wish to thank M.L.~Mangano for making the squared amplitude for $W$
production of Ref.~\cite{Mangano:1990gs} available to us for comparison,
Carl Schmidt for useful discussions, and Tim Stelzer for his valuable help
with MADGRAPH. JRA acknowledges the financial support
of The Danish Research Agency.

\appendix

\section{Multiparton kinematics}
\label{sec:appa}

We consider the production of  a $W (\to e \nu )$ and two
jets of momenta $p_{a'}$ and $p_{b'}$,
in the scattering between two partons of
momenta $p_a$ and $p_b$ ($p_a^0<0$ and $p_b^0<0$).\footnote{Conventionally,
in the helicity amplitudes all momenta are always taken as outgoing.
Partons which in a physical channel are incoming are then identified by
the (negative) sign of their energy.}

Using light-cone coordinates $p^{\pm}= p_0\pm p_z $, and
complex transverse coordinates $p_{\perp} = p^x + i p^y$, with scalar
product $2 p\cdot q = p^+q^- + p^-q^+ - p_{\perp} q^*_{\perp} - p^*_{\perp}
q_{\perp}$, the 4-momenta are,
\begin{eqnarray}
p_a &=& \left(p_a^+/2, 0, 0,p_a^+/2 \right)
     =  \left(p_a^+ , 0; 0, 0 \right)\, ,\nonumber \\
p_b &=& \left(p_b^-/2, 0, 0,-p_b^-/2 \right)
     =  \left(0, p_b^-; 0, 0\right)\, ,\label{in}\\
p_i &=& \left( (p_i^+ + p_i^- )/2,
                {\rm Re}[p_{i_\perp}],
                {\rm Im}[p_{i_\perp}],
                (p_i^+ - p_i^- )/2 \right)\nonumber\\
    &=& \left(|p_{i_\perp}| e^{y_i}, |p_{i_\perp}| e^{-y_i};
|p_{i_\perp}|\cos{\phi_i}, |p_{i_\perp}|\sin{\phi_i}\right)\, \,,
\;i=3(b'),4(a'),5(\bar\ell),6(\ell) \nonumber\,
\end{eqnarray}
where the first notation is the standard representation
$p^\mu =(p^0,p^x,p^y,p^z)$, while in the second we have the + and -
components to the left of the semicolon ,
and to the right the transverse components.
$y$ is the parton rapidity and $\phi$ is the azimuthal angle between the
vector $p_{\perp}$ and an arbitrary vector in the transverse plane.
From the momentum conservation ($i=b',a',{\bar\ell},\ell$),
\begin{eqnarray}
0 &=& \sum_{i=3}^{6} p_{i_\perp}\, ,\nonumber \\
p_a^+ &=& -\sum_{i=3}^{6} p_i^+\, ,\label{nkin}\\
p_b^- &=& -\sum_{i=3}^{6} p_i^-\, ,\nonumber
\end{eqnarray}
the Mandelstam invariants may be written as,
\begin{eqnarray}
\hat s_{ij} &=& 2 p_i\cdot p_j = p_i^+ p_j^- + p_i^- p_j^+
- p_{i_\perp} p_{j_\perp}^* - p_{i_\perp}^* p_{j_\perp}\, .\nonumber
\end{eqnarray}
so that
\begin{eqnarray}
\hat s &=& 2 p_a\cdot p_b = \sum_{i,j=3}^{6} p_i^+ p_j^- \,,\nonumber\\
\hat s_{ai} &=& 2 p_a\cdot p_i = -\sum_{j=3}^{6} p_i^- p_j^+ \,,\label{inv}\\
\hat s_{bi} &=& 2 p_b\cdot p_i = -\sum_{j=3}^{6} p_i^+ p_j^- \nonumber.
\end{eqnarray}

The spinor products are defined as
\begin{eqnarray}
\langle p_i - | p_j + \rangle &\equiv& \langle i j\rangle \,,\nonumber\\
\langle p_i + | p_j - \rangle &\equiv& \left[ i j \right] \,,\label{spipro}\\
\langle p_i-| \slash  \!\!\! p_k  |p_j-\rangle &\equiv& \langle i | k |
j\rangle \,.\nonumber
\end{eqnarray}
Using the above spinor representation, the spinor products for the
momenta (\ref{in}) are
\begin{eqnarray}
\langle p_i p_j\rangle &=& p_{i_\perp}\sqrt{p_j^+\over p_i^+} - p_{j_\perp}
\sqrt{p_i^+\over p_j^+}\, , \nonumber\\
\langle p_a p_i\rangle &=& - i \sqrt{-p_1^+
\over p_i^+}\, p_{i_\perp}\, ,\label{spro}\\
\langle p_i p_b\rangle &=&
i \sqrt{-p_b^- p_i^+}\, ,\nonumber\\
\langle p_a p_b\rangle
&=& -\sqrt{\hat s}\, ,\nonumber
\end{eqnarray}
where we have used the mass-shell condition
$|p_{i_\perp}|^2 = p_i^+ p_i^-$.
The spinor products fulfill the identities ($ i\equiv  p_i, j\equiv p_j$),
\begin{eqnarray}
\langle i j\rangle &=& - \langle j i\rangle \,,\nonumber\\
           \left[ i j \right] &=& - \left[ j i \right]\,,\nonumber \\
\langle i j\rangle^* &=& {\rm sign}(p^0_i p^0_j) \left[ j i\right]\,,
\label{flips4}\nonumber\\
\left( \langle i+| \gamma^{\mu}  |j+\rangle \right)^* &=&
{\rm sign}(p^0_i p^0_j)
\langle j+| \gamma^{\mu}  |i+\rangle \label{flipc2}, \\
\langle i j \rangle \left[ji\right] &=&
2p_i\cdot p_j = \hat{s}_{ij}\, ,\nonumber \\
\langle i+| \slash \!\!\! k  |j+\rangle  &=&
\left[ i k \right] \langle k j \rangle ,\nonumber \\
\langle i-| \slash  \!\!\! k  |j-\rangle  &=&
\langle i k \rangle \left[k j\right]\,.\nonumber
\end{eqnarray}
%

\section{Next-to-leading corrections in the forward-rapidity region}
\label{sec:appc}

We consider the production of particles $p_a'$, $p_e$, $p_{\nu}$
in the forward-rapidity region of parton $p_a$,
\begin{equation}
y_{q} \simeq y_{e} \simeq y_{\nu} \gg y_{b'};\qquad
|p_{e_\perp}| \simeq |p_{\nu_\perp}|
\simeq |p_{q_\perp}| \simeq |p_{b'_\perp}|\, ,\label{qmrapp}
\end{equation}
Momentum conservation (\ref{nkin}) simply generalizes to,
\begin{eqnarray}
p_a^+ &\simeq& -(p_q^+ + p_e^+ + p_\nu^+)\, ,\label{frkapp}\\
p_b^- &\simeq& -p_{b'}^-\, .
\end{eqnarray}
and accordingly the Mandelstam invariants (\ref{inv}) may be written as,
\begin{eqnarray}
\hat s &=& 2 p_a\cdot p_b\, \simeq\,
(p_q^+ + p_e^+ + p_\nu^+) p_{b'}^-\, ,\nonumber \\
\hat u &=& 2 p_a\cdot p_{b'}\, \simeq\, - (p_q^+ + p_e^+ + p_\nu^+) p_{b'}^-\,
,\nonumber \\
\hat u_k &=& 2 p_b\cdot p_k\, \simeq\, - p_k^+ p_{b'}^-\, ,\nonumber
\qquad k=q,e,\nu\\
\hat t_k &=& 2 p_a\cdot p_k\,\simeq\, - (p_q^+ + p_e^+ + p_\nu^+) p_k^-\, ,
\qquad k=q,e,\nu\label{frinv}\\
\hat t &=& 2 p_b \cdot p_{b'}\, \simeq\, - |p_{b'_\perp}|^2\, ,\nonumber
\end{eqnarray}
to leading accuracy.
The spinor products (\ref{spro}) become
\begin{eqnarray}
\langle p_a p_b\rangle &=&
-\sqrt{\hat s} \simeq - \sqrt{(p_q^+ + p_e^+ + p_\nu^+) p_{b'}^-}\,
,\nonumber\\
\langle p_a p_{b'}\rangle &=&
-i \sqrt{-p_a^+\over p_{b'}^+}\, p_{b'_\perp} \simeq i
{p_{b'_\perp}\over |p_{b'_\perp}|} \langle p_a p_b\rangle\, ,\nonumber\\
\langle p_a p_k\rangle &=& -i \sqrt{-p_a^+\over p_k^+}\, p_{k_\perp}
\simeq -i \sqrt{p_q^+ + p_e^+ + p_\nu^+ \over p_k^+} p_{k_\perp}\, ,
\nonumber \qquad \qquad k=q,e,\nu\\
\langle p_k p_b\rangle &=& i \sqrt{-p_b^- p_k^+}\,
\simeq i \sqrt{p_k^+ p_{b'}^-}\, ,\label{frpro}\qquad k=q,e,\nu\nonumber\\
\langle p_{b'} p_b\rangle &=& i \sqrt{-p_b^- p_{b'}^+}\,
\simeq i |p_{b'_\perp}|\, ,\nonumber\\
\langle p_k p_{b'}\rangle &=&
p_{k_\perp}\sqrt{p_{b'}^+\over p_k^+} - p_{b'_\perp}
\sqrt{p_k^+\over p_{b'}^+} \simeq - p_{b'_\perp}\,
\sqrt{p_k^+\over p_{b'}^+}\, ,\nonumber \qquad k=q,e,\nu \,.
\end{eqnarray}

\vskip .2 cm


\def\Journal#1#2#3#4{{#1} {\bf #2}, #3 (#4)}

\def\NCA{\em Nuovo Cimento}
\def\NIM{\em Nucl. Instrum. Methods}
\def\NIMA{{\em Nucl. Instrum. Methods} A}
\def\NPB{{\em Nucl. Phys.} B}
\def\PLB{{\em Phys. Lett.}  B}
\def\PRL{\em Phys. Rev. Lett.}
\def\PRD{{\em Phys. Rev.} D}
\def\ZPC{{\em Z. Phys.} C}

\end{document}